\documentclass{article}

\usepackage{PRIMEarxiv}

\usepackage[utf8]{inputenc} 
\usepackage[T1]{fontenc}    
\usepackage{hyperref}       
\usepackage{url}            
\usepackage{booktabs}       
\usepackage{amsfonts}       
\usepackage{nicefrac}       
\usepackage{microtype}      
\usepackage{lipsum}
\usepackage{fancyhdr}       
\usepackage{graphicx}       
\graphicspath{{media/}}     

\PassOptionsToPackage{table, x11names}{xcolor}
\usepackage[table,x11names]{xcolor}
\usepackage[utf8]{inputenc}
\usepackage[T1]{fontenc} 
\usepackage{xcolor}
\usepackage{multirow}
\usepackage{caption}
\usepackage{subcaption}
\usepackage{comment}
\usepackage{mdframed}

\graphicspath{{media/}}   
\definecolor{mydarkblue}{HTML}{073b4c}
\definecolor{plgreen}{HTML}{CDFADB}
\definecolor{plyellow}{HTML}{F6FDC3}
\definecolor{plorange}{HTML}{FFCF96}
\definecolor{plred}{HTML}{FF8080}

\definecolor{ResNetCol}{HTML}{000004}
\definecolor{EfficientNetCol}{HTML}{bc3754}
\definecolor{SwinTransformerCol}{HTML}{57106e}
\definecolor{MalConvCol}{HTML}{f98e09}
\definecolor{LightGBMCol}{HTML}{ffff00}

\usepackage{pgfplots}
\pgfplotsset{compat=1.18}
\newcommand{\fillbox}[1]{
  \colorbox{#1}{\rule{0pt}{8pt}\rule{8pt}{0pt}}\;
}

\title{Assessing the Impact of Packing on Machine Learning-Based Malware Detection and Classification Systems
\thanks{\textit{\underline{Citation}}: 
\textbf{Daniel Gibert, Nikolaos Totosis, Constantinos Patsakis, Giulio Zizzo, Quan Le. Assessing the Impact of Packing on Machine Learning-Based Malware Detection and Classification Systems. Pages.... DOI:000000/11111.}} 
}

\author{
  Daniel Gibert\\
  CeADAR, University College Dublin\\
  Dublin, Ireland \\
  \And
  Nikolaos Totosis \\
  University of Piraeus \\
  Athens, Greece \\
  \And
  Constantinos Patsakis\\
  Athena Research Center, University of Piraeus\\
  Athens, Greece \\
  \And
  Giulio Zizo \\
  IBM Research Europe\\
  Dublin, Ireland \\
  \And
  Quan Le \\
  CeADAR, University College Dublin\\
  Dublin, Ireland \\
}

\begin{document}
\maketitle

\begin{abstract}
The proliferation of malware, particularly through the use of packing, presents a significant challenge to static analysis and signature-based malware detection techniques. The application of packing to the original executable code renders extracting meaningful features and signatures challenging. To deal with the increasing amount of malware in the wild, researchers and anti-malware companies started harnessing machine learning capabilities with very promising results. However, little is known about the effects of packing on static machine learning-based malware detection and classification systems. This work addresses this gap by investigating the impact of packing on the performance of static machine learning-based models used for malware detection and classification, with a particular focus on those using visualisation techniques. To this end, we present a comprehensive analysis of various packing techniques and their effects on the performance of machine learning-based detectors and classifiers. Our findings highlight the limitations of current static detection and classification systems and underscore the need to be proactive to effectively counteract the evolving tactics of malware authors.
\end{abstract}


\section{Introduction}
The generation of continually changing products and services in the cyber domain has led to a large opportunity space for capital through cybercrime. These developments are very attractive for criminals, as they significantly increase their outreach while also allowing for automation. 
This trend maps to the number of malware samples per year, which in the past decade has stabilised to more than 80 million \cite{avatlas}. 
This massive number of samples, along with the increase in sophistication and impact of malware, requires better and more advanced countermeasures \cite{malware}. To fill this gap, many researchers and anti-malware companies are harnessing artificial intelligence (AI) and machine learning (ML).

Of particular interest in this work are the strong results reported by using machine learning and artificial intelligence to classify binaries as benign or malicious, or to perform multiclass malware family classification using visualization techniques. Research in this domain has generated numerous articles in recent years, most of which report almost perfect results with very high accuracy and precision. Querying Scopus for relevant publications, 
resulted in well beyond 400 documents from 2008 to 2024, 
clearly showing the research interest in this research field.

\begin{figure}[!th]
    \centering
    \begin{subfigure}{.48\columnwidth}
        \centering
        \includegraphics[width=.35\columnwidth]{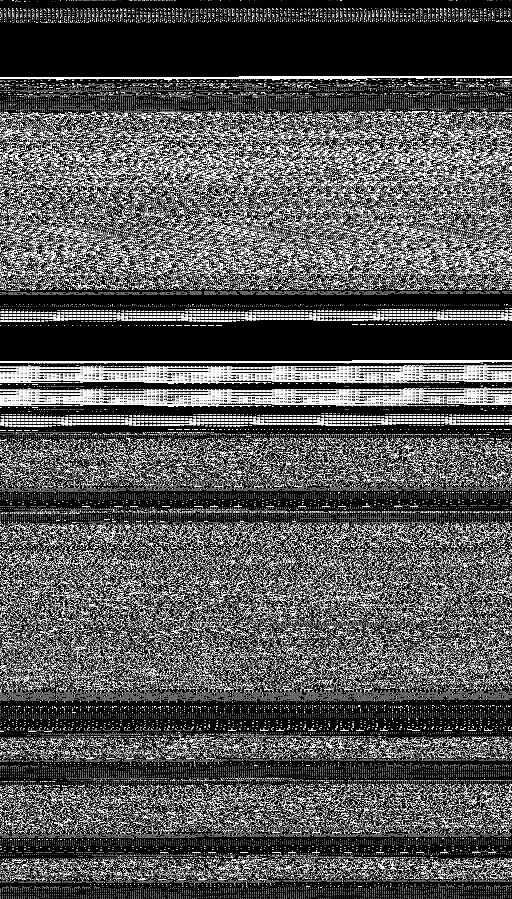}
        \includegraphics[width=.35\columnwidth]{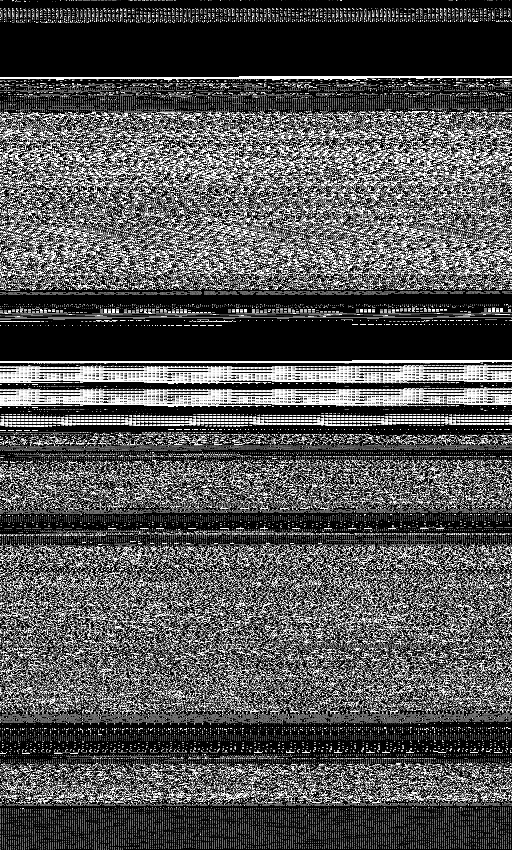}
        \caption{VB.AT samples}
    \end{subfigure}
    \begin{subfigure}{.45\columnwidth}
        \centering
        \includegraphics[width=.485\columnwidth]{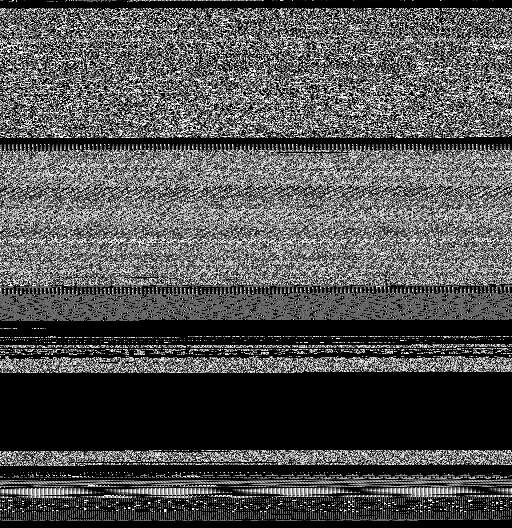}
        \includegraphics[width=.485\columnwidth]{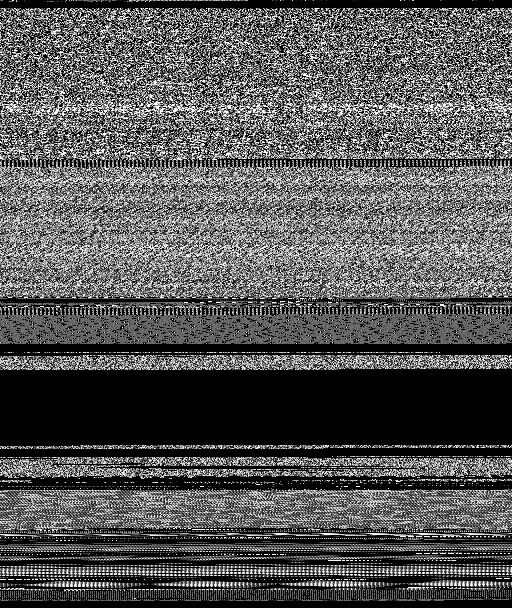}
        \caption{Swizzor.gen!I samples}
    \end{subfigure}
    \caption{Sample images from the \texttt{Malimg} dataset with the visualisation of binaries from two malware families. }
    \label{fig:malimg_images}
\end{figure}

The core concept of this line of research is to convert binaries into images and then try to find common patterns in the images to classify them accordingly, something that machine learning methods have shown to do exceptionally well for many years. Indeed, as observed in Figure \ref{fig:malimg_images} that contains samples from a reference dataset of this domain \cite{10.1145/2016904.2016908}, the visual representations of malware exhibit obvious visual similarities between them, even if the files are different. Thus, the conceptual approach sounds very promising, and the results are also in favor. However, is this enough to actually state that these methods are accurate and that they illustrate the provenance of machine learning and artificial intelligence in malware detection and classification?

Since generating so many malware samples manually is impossible, malware authors use various methods to create variants of their binaries. However, the tooling often used to generate them and hide their payloads, e.g., packers, leaves an identifiable footprint. Traditionally, malware analysts would use tools such as YARA rules to identify the packer and the underlying malware. Therefore, it is natural to consider whether the visual characteristics depend on the tooling and whether this can be used to bypass such detectors.    
The above leads us to formulate two main research questions that we try to answer in this work:
\begin{enumerate}
    \item Does a bias in the distribution of packers used in benign and malicious executables cause malware detectors to learn "benign" and "malicious" packing routines? How does a bias in the distribution of packers affect malware classifiers?
    \item Do packers limit the applicability of static machine learning-based malware detectors and classifiers? What effort is required to bypass detection or raise false flags in malware detectors and cause a misclassification in malware classifiers?
\end{enumerate}

To address these questions, our work makes the following contributions:
\begin{itemize}
    \item We conduct an extensive analysis of the bias in the distribution of packers used in benign and malicious executables, demonstrating how this bias influences the learning process of malware detectors. Our findings reveal that malware detectors can indeed develop a reliance on specific packing routines as indicators of benign or malicious behaviour, or among malware families, which can significantly impact their performance and reliability.
    \item We provide a detailed study of the performance of static machine learning-based malware detectors and classifiers, including feature-based detectors, end-to-end detectors, and greyscale image-based detectors, when subjected to packing techniques, and we identify scenarios where packers successfully bypass detection or trigger false positives.
    \item We examine how 8 different packers and obfuscation tools affect the performance of static machine learning-based malware detectors and classifiers, including well-known executable file compressors such as UPX and MPress, crypters such as Hyperion, protectors such as Enigma and Themida, and obfuscation tools such as Mangle. Through a series of experiments, we demonstrate which packing techniques are more effective against each type of detector, highlighting the limitations of static detectors against obfuscated executables.
\end{itemize}

The rest of this work is structured as follows. In the next section, we provide an overview of the related work. Then, in Section~\ref{sec:analysis}, we perform a thorough analysis of two reference datasets in the field, which motivates us to dig deeper into this research, as the analysis of the underlying data justifies many visual characteristics. In Section \ref{sec:methodology}, we establish our methodology, which is used afterwards in Section \ref{sec:experiments}, where we detail our experiments and results. Finally, the article concludes by summarizing our contributions and discussing ideas for future work.

\section{Related Work}
\label{sec:related}
Machine learning has increasingly been adopted to detect and classify malware due to its ability to learn patterns and characteristics from large amounts of data. Despite showing promising results in the detection and classification of binaries, ML-based solutions are vulnerable against adversarial examples~\cite{DBLP:journals/tifs/DemetrioBLRA21,DBLP:journals/tissec/DemetrioCBLAR21}. In addition, previous studies on the limits of ML-based models based on static features against packing have highlighted their limitations \cite{aghakhani2020malware}. This study builds upon their work, and it studies the impact of packers and obfuscation tools on various types of static ML-based malware detection systems, including feature-based detectors, end-to-end detectors, and greyscale image-based detectors, showcasing the limitations of state-of-the-art static ML-based detectors and classifiers.

Feature-based detectors rely on feature engineering and expert knowledge to extract specific features or characteristics from binary files that are believed to be indicative of malware. The most well-known feature-based detector is the EMBER LightGBM model \cite{2018arXiv180404637A}. This detector refers to a LightGBM model trained using various features to represent a computer program, including general file information (e.g., file size, PE header data), header information (e.g., COFF and Optional headers), imported and exported functions (from Import and Export Address Tables), section information (e.g., section size and entropy), byte histograms, byte-entropy histograms (computed using a sliding window), and string information (e.g., statistics about printable strings).

On the other hand, end-to-end malware detectors refer to deep learning models (e.g., convolutional neural networks) that take raw binary files as input and output a classification (benign or malicious) without human intervention. End-to-end detectors eliminate the need to manually extract expert-crafted features, automating the feature learning and classification process. A well-known end-to-end detector is MalConv~\cite{DBLP:conf/aaai/RaffBSBCN18}, a shallow convolutional neural network that consists of a gated convolutional layer with large filters, followed by a global max pooling layer and a fully connected layer. 

Finally, greyscale image-based detectors take the raw binary data of a file and convert them into a greyscale image~\cite{10.1145/2016904.2016908}. Each byte in the binary is mapped to a pixel in the image, with the byte value (0-255) corresponding to the greyscale intensity of the pixel. Then, the resulting image is resized, normalised, and classified using a convolutional neural network (CNN)~\cite{DBLP:journals/virology/GibertMPV19} such as ResNet~\cite{DBLP:conf/cvpr/HeZRS16} and EfficientNet~\cite{DBLP:conf/icml/TanL21}, or with vision transformer models~\cite{DBLP:conf/iclr/DosovitskiyB0WZ21,9710580}. The greyscale image representation of malware has been mainly used to classify malware into families because, as shown in Figure \ref{fig:malimg_images}, the visual representation of malware exhibits obvious visual similarities between samples belonging to the same family while being different from those samples belonging to a different family.

\section{Motivation and dataset analysis}
\label{sec:analysis}
Setting aside the efficacy of different malware detectors using visualisations of malware binaries, it is essential to understand why they are efficient. However, this cannot be answered by using explainability frameworks, e.g., Shapley Additive Explanations (SHAP) \cite{merrick2020explanation} or Local Interpretable Model-Agnostic Explanations (LIME) \cite{ribeiro2018semantically}, as the answers will refer to visual data. We argue that before going through the visual features and their weights, one has to examine the underlying data, which, in this case, is the malware itself. 

The most principled way is to directly and manually examine one of the reference datasets in this domain, with the most broadly used being the \texttt{Malimg} dataset \cite{10.1145/2016904.2016908}. The dataset dates back to 2011 and contains 9,339 malware samples belonging to 25 malware families and consists of visualisations of these binaries into greyscale images. Unfortunately, the image conversion that has been used to construct the dataset is lossy, as some bytes at the end of the files are removed to fit them to a specific image width. Thus, it is not possible to directly derive all the original binaries from the images of the dataset. Since each image is named after the hash of the original binary, one can easily check that by reversing the process, only 3666 of the samples can be extracted. Moreover, the names can be used to retrieve intelligence reports from various OSINT services (e.g., VirusTotal, Abuse.ch, and VX Underground) or even reconstruct them if only a few bytes are missing, e.g., by brute forcing some bytes at the end of the file to find the suffix to match the hash of the name. 

Using the above methods, we managed to reconstruct and collect intelligence reports from 8263 files (87.36\% of the original files). For the rest of the 1076 files, we have the truncated version of these files, so some information is missing, and there is no available information from OSINT sources. The information per family is shown in Table \ref{tbl:malimg}. Note that information such as the names of the sections, the hashes of the sections that have not been tampered with by the chopping, the import tables, etc., can be safely extracted from these files. Evidently, the fuzzy hashes (ssdeep and TLSH) cannot be properly computed on the truncated files. Thus, we have all the information from the original 8,263 files and partial information from 1,076 files. 

\begin{table}[th]
    \centering
    \scriptsize
    \caption{A breakdown of the \texttt{Malimg} dataset in families as per the original and the retrieved samples. Rows highlighted in \fillbox{plgreen} denote malware families which are individually distinguished. Rows highlighted in \fillbox{plyellow} denote malware families which are distinguished from others as part of a group of two or more families. Rows highlighted in \fillbox{plred} denote malware families whose most samples are distinguished by the packer/compiler.}
    \label{tbl:malimg}
    \begin{tabular}{lrr}
    \toprule
    \textbf{Family} & \textbf{Original Samples}  & \textbf{Retrieved Intelligence}\\
    \midrule
\rowcolor{plgreen}Adialer.C	&	122	&	22	\\
\rowcolor{plgreen}Agent.FYI	&	116	&	116	\\
\rowcolor{plyellow}Allaple.A	&	2949	&	2818	\\
\rowcolor{plyellow}Allaple.L	&	1591	&	1570	\\
\rowcolor{plgreen}Alueron.gen!J	&	198	&	193	\\
\rowcolor{plgreen}Autorun.K	&	106	&	105	\\
C2LOP.gen!g	&	200	&	166	\\
C2LOP.P	&	146	&	144	\\
\rowcolor{plgreen}Dialplatform.B	&	177	&	177	\\
\rowcolor{plgreen}Dontovo.A	&	162	&	162	\\
\rowcolor{plred}Fakerean	&	381	&	321	\\
\rowcolor{plgreen}Instantaccess	&	431	&	52	\\
\rowcolor{plyellow}Lolyda.AA1	&	213	&	213	\\
\rowcolor{plyellow}Lolyda.AA2	&	184	&	184	\\
\rowcolor{plyellow}Lolyda.AA3	&	123	&	123	\\
\rowcolor{plgreen}Lolyda.AT	&	159	&	156	\\
\rowcolor{plgreen}Malex.gen!J	&	136	&	17	\\
\rowcolor{plgreen}Obfuscator.AD	&	142	&	16	\\
\rowcolor{plred}Rbot!gen	&	158	&	153	\\
\rowcolor{plgreen}Skintrim.N	&	80	&	80	\\
Swizzor.gen!E	&	128	&	126	\\
Swizzor.gen!I	&	132	&	132	\\
\rowcolor{plgreen}VB.AT	&	408	&	326	\\
\rowcolor{plgreen}Wintrim.BX	&	97	&	94	\\
\rowcolor{plgreen}Yuner.A	&	800	&	797	\\ \midrule
\textbf{Total}	&	\textbf{9339}	&	\textbf{8263}	\\

    \bottomrule\\
    
    \end{tabular}
\end{table}

\begin{figure*}[th]
\centering
  \includegraphics[width=\textwidth]{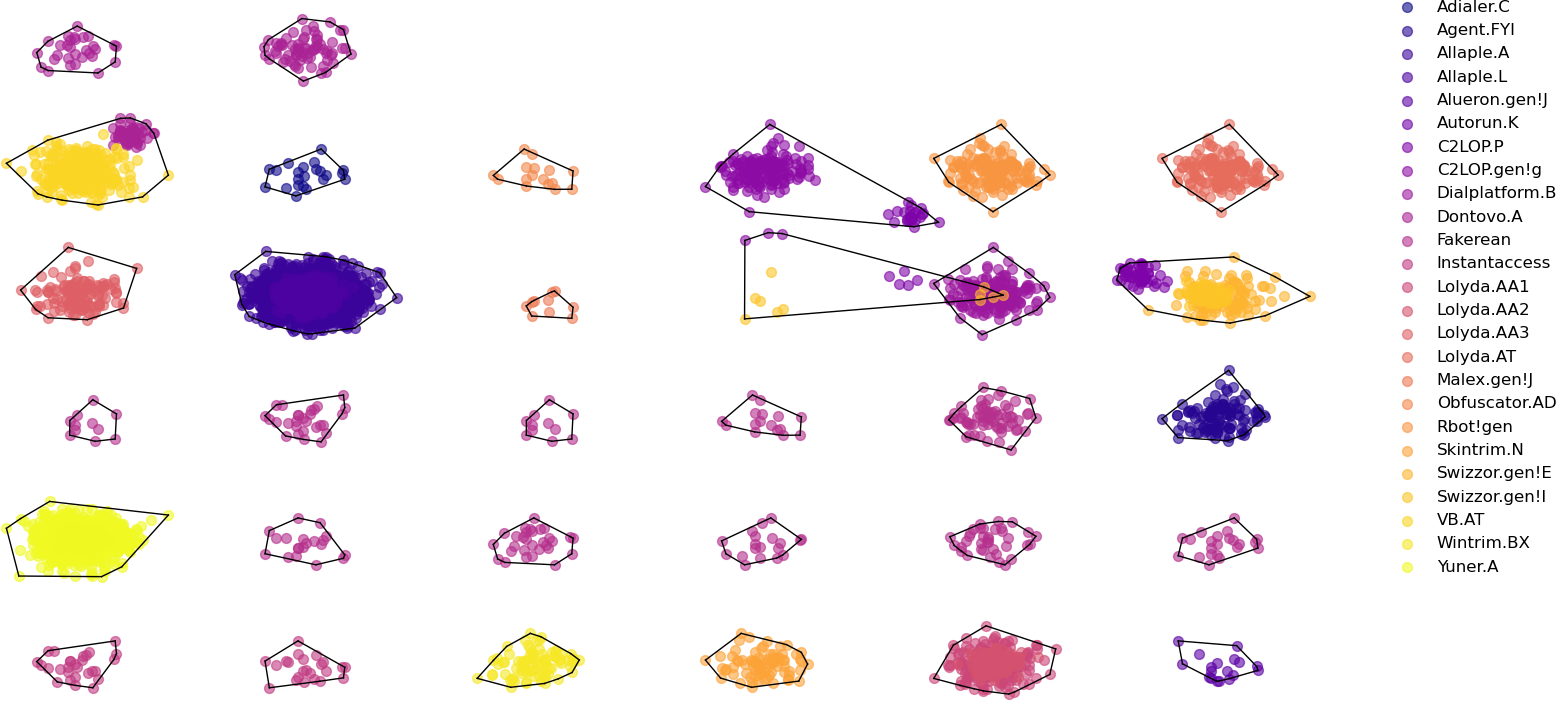}
  \caption{\texttt{Malimg} dataset clustered by TLSH and ssdeep, displaying only clusters with more than 10 items.}
  \label{fig:cluster}
\end{figure*}

Based on the collected data, we considered it essential to examine the structural data of the files. First, we must note that each file in our dataset is a PE32 executable. Moreover, for each file, we have two fuzzy hashes that are widely used, ssdeep \cite{kornblum2006identifying} and TLSH \cite{oliver2013tlsh}. These two hashes allow for fast clustering of similar files. Using these hashes, we cluster the data as shown in Figure \ref{fig:cluster}. From the original 8263 files of the dataset, by only using the ssdeep similarity score above 70 and a TLSH distance of less than 100, we can see that a significant part of the dataset is correctly clustered, clearly indicating that the files of each family are similar at the byte level. The latter clearly implies that the images that would be extracted from similar files would also be similar. 

However, there are further structural correlations. More precisely, PE32 files are split into sections, building blocks that contain actual program data and resources. Each section has a specific name and characteristics, with the most common sections being:
\begin{itemize}
    \item \texttt{.text}: Containing the executable code (instructions) for the program.
    \item \texttt{.rdata}: Containing read-only data, such as strings and constants.
    \item \texttt{.data}: Containing initialized data variables.
    \item \texttt{.rsrc}: Containing resources such as icons, menus, and images.
    \item \texttt{.reloc}: An optional section containing relocation information to adjust addresses when loading the file.
\end{itemize}
Nevertheless, the number of sections, their size and name depend on the compiler, linker, builder, and packer that has been used to generate the executable.

Taking the hash of each section, we notice many files from specific families containing the exact same sections. Therefore, even if there are differences in other sections of the executables, at the byte level, some sections are identical. Again, we can safely assume that these byte-level similarities lead to visual similarities once the binaries are converted to images. Indicatively, all samples of the VB.AT family have a section whose MD5 hash is \texttt{30695b8f3e042a947d4aa46b7f80da27}, while Yuner samples have a section with MD5 hash being \sloppy{\texttt{beafbde081a00045c5646597f1b5b055}}. Moreover, by examining the section names, it is easy to find all samples of the Alueron family as they have sections with names \texttt{weij} (in various upper/lowercase variations), \texttt{PAGELK}, \texttt{CODE}, or \texttt{itext}. Similarly, by pruning the above, all samples from the Lolyda.AA* families are the only ones that contain a \texttt{.reloc} section. On the contrary, the absence of the \texttt{.rsrc} section can be checked to distinguish samples from the two Allaple.* families. 

Combining the above, one can understand that the structural differences among the families of this dataset are very distinctive at the byte level. Indeed, in Table \ref{tbl:malimg}, we have highlighted the families that can be uniquely identified in green and the families of Lolyda.AA* can be immediately distinguished from the remaining ones with yellow. Therefore, 14 of the 25 malware families can be effortlessly classified with absolute accuracy and precision, while three more can be easily distinguished. 

The remaining families have other unique characteristics that can be used to cluster. For instance, using peid\footnote{\url{https://github.com/packing-box/peid}}, one can observe other attributes of the binaries. Indicatively, the UPX packer is only used by almost all (158/159) samples of the Lolyda.AT family. Most samples of the Fakerean family (307/381) are compiled using Borland Delphi. Similarly, FSG has been used by most Rbot!gen samples (135/158).

\begin{figure*}[th]
\centering
\includegraphics[width=\textwidth]{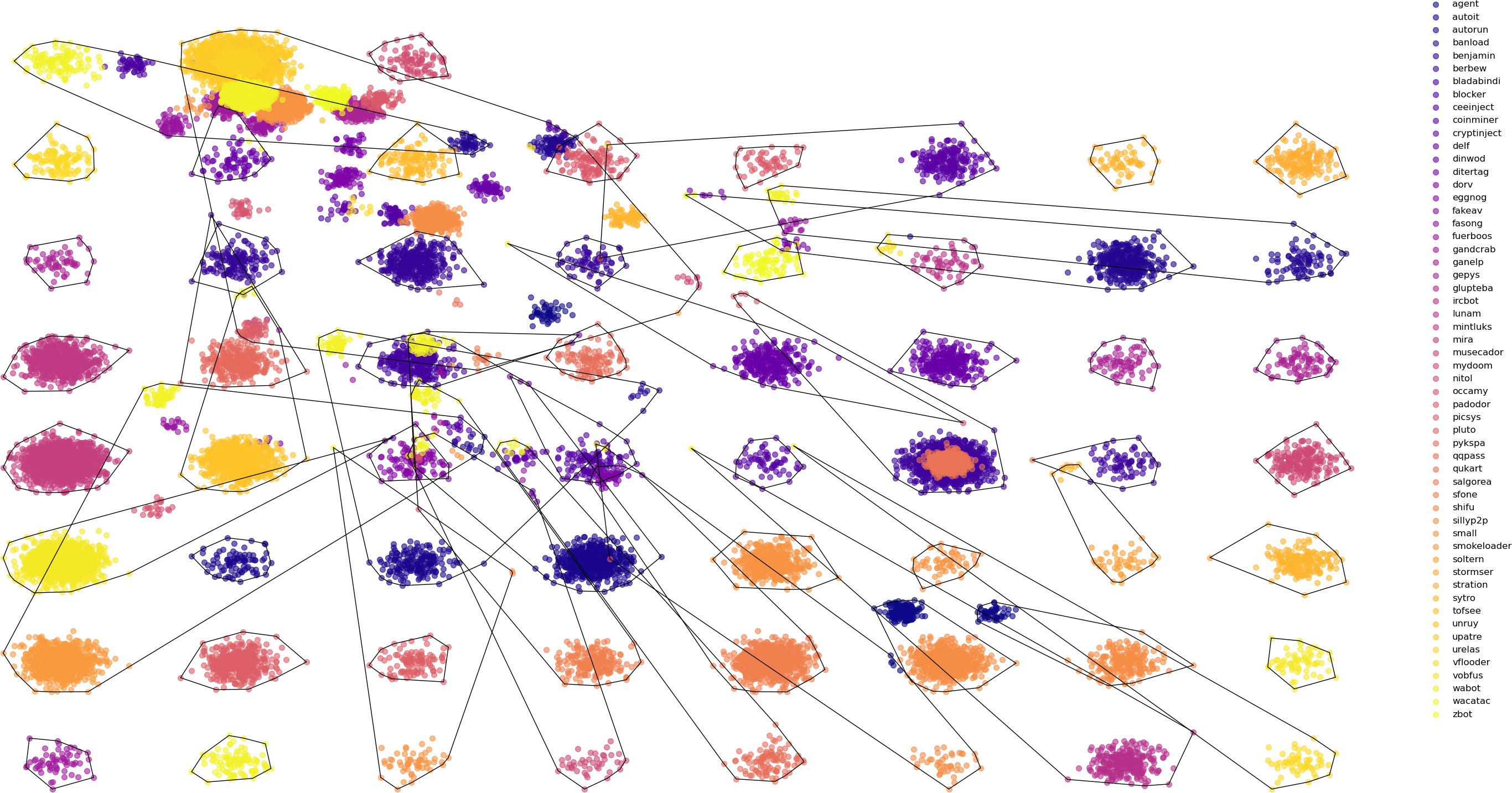}
  \caption{
  \texttt{BODMAS} dataset clustered by TLSH and ssdeep, displaying only clusters with more than 10 items.}
  \label{fig:bodmas_clusters}
\end{figure*}

We performed the same analysis on the \texttt{BODMAS} dataset \cite{DBLP:conf/sp/YangCLA021}. Again, we noticed many structural correlations, as in the case of the \texttt{Malimg} dataset. For instance, from the 920 samples of the drolnux family, 909 have a section with the hash \texttt{3a99bef55e205511b4925c188c55c719}, 508 of the 509 samples of the fakeav family share a section with the hash \texttt{c29546ca3cc0483df253a231a85fa09f} and all the samples (1071) of benjamin family share a section with hash \texttt{07f444205f1a50b292507ffabe5d505f}. Similarly, 1018 of the 1054 samples of the musecador family have \texttt{d2bf2bc66c5e49a85254cd29b19046bd} as imphash, and all the samples of the station family (158) have the same imphash \texttt{547cd05356c429dc57b17bf0fd6daf12}. However, for the sake of brevity, we do not provide complete clustering. The clustering of the dataset using only TLSH and ssdeep for clusters with a cardinality of more than 10 is illustrated in Figure \ref{fig:bodmas_clusters}.    

From our analysis, it is clear that, when examining both datasets, there are obvious structural issues that do not require visual classification to be identified. In fact, these features can be easily collected and used to classify malware accurately. Moreover, we observed that some features may belong to multiple families. For instance, an \texttt{.rdata} section with hash \texttt{615bb796c5e715fc753114d5343fa5b4} is shared by three families in the \texttt{BODMAS} dataset, namely sillyp2p (877), small (152) and agent (233). So, practically, we have three different families using exactly the same section. Similarly, we can find that the \texttt{.idata} section with hash \texttt{708cff90e55fcc1f43ce49fc7ad6f7f4} is used by berbew (1663), qukart (821) and padodor (5). Given the modus operandi of modern malware \cite{patsakis2024malware}, the above can be attributed to using the same tooling by malware authors to protect their binaries. Thus, we can safely assume that we have the same packer or obfuscator.  

When considering the latter with the fact that visual malware classifiers are looking for image similarities, it is natural to ponder whether they classify malware or the obfuscators and packers the malware uses. Evidently, this is a completely different problem. Moreover, this opens the door to creating numerous adversarial samples that bypass these detectors by, e.g., using a new packer, raising false flags \cite{skopik2020under,bartholomew2016wave}, or even having many false negatives, should legitimate software use abused packers for protecting their intellectual property rights; see the case of the Themida packer.  

The above structural features justify the visual similarities that CNNs and other such methods exploit. Given the efficacy of the aforementioned structural features, how easily they can be extracted, the efficacy of YARA rules, etc., it is obvious to question the dependency on AI and machine learning with visual features for such tasks. The latter becomes more relevant because of the need to have an ample number of labelled samples to train them. On the contrary, with a handful of samples, the structural similarities can be extracted and used with far fewer computational resources. Moreover, the latter observations suggest the heavy reliance of malware classifiers with visual input on packers. We anticipate that using packers can significantly impact the classification of malware samples.

\section{Methodology}
\label{sec:methodology}
In this study, we explore the impact of packing on the performance of static ML-based malware detectors and classifiers. While malware detectors aim to identify malicious software, malware classifiers categorise them into specific malware families. We hypothesise that packing and obfuscation techniques substantially reduce the effectiveness of these ML-based systems, while the extent of their impact likely varies based on the diversity of packers and the number and type of packed executables encountered during the training phase. To help answer our hypothesis, we pose the following research questions and conduct a series of experiments specifically designed to shed light on how packing influences both the detection and classification capabilities of these systems, particularly under different training conditions.

\begin{mdframed}[backgroundcolor=gray!10,linewidth=0.5pt]
    \noindent \textbf{Research Question 1: }Does a bias in the distribution of packers used in benign and malicious executables cause malware detectors to learn "benign" and "malicious" packing routines? How does a bias in the distribution of packers affect malware classifiers?
\end{mdframed}

There are several reasons that could lead to biases in the distribution of packers used in benign and malicious executables. For instance, the period during which the data were collected affects the representation of the packing techniques in the samples since newer packing and obfuscation techniques have been developed, and older ones have become obsolete. Thus, a model trained with samples packed using obsolete techniques may perform poorly, on newer, unseen, packing techniques. The time perspective has been discussed in \cite{wadkar2020detecting,ceschin2023fast} but focuses on the malware family perspective and concept drift. Likewise, the capability of existing tools to detect and unpack certain types of packed executables might influence the prevalence of these packers in the datasets. Even more, the intent behind the software affects the packer's choice. Malicious software may prefer to use specific packers that are more effective at evading detection, whereas benign software may prefer packers simply for compression or to protect intellectual property.

The bias in the distribution of packers may impact the performance of malware detectors and classifiers in unexpected ways. In the context of malware detection, a bias in the distribution of packers used in benign and malicious executables may cause detectors to learn \emph{benign} and \emph{malicious} packing routines. In the context of malware classification, using packing by malware families may cause the classifier to learn to classify packers instead of malware or to misclassify packed malware as belonging to specific malware families. Thus we can pose the following research question:

\begin{mdframed}[backgroundcolor=gray!10,linewidth=0.5pt]
    \noindent \textbf{Research Question 2: } Do packers limit the applicability of static machine learning-based malware detectors and classifiers? What effort is required to bypass detection or raise false flags in malware detectors and cause a misclassification in malware classifiers?
\end{mdframed}

Packers obfuscate the content of executables through various techniques aimed at making the executable's code and data less readable and harder to analyse. Consequently, we hypothesise that ML-based detectors and classifiers, particularly those relying on deep learning, may struggle to differentiate between benign and malicious executables that have been packed with the same packer or to distinguish between malware families whose executables have been packed with the same packer. As a result, such detection systems may be prone to generating an alarming rate of false positives on packed goodware, while attackers could exploit this weakness to evade detection by disguising malware using common packers.

\subsection{Packers}
\label{sec:packers}
The following packers and tools have been executed either using wine\footnote{\url{https://www.winehq.org/}} on a Ubuntu system or using the ProtectMyTooling tool\footnote{\url{https://github.com/mgeeky/ProtectMyTooling}} on a Virtual Machine running Windows 10.

\subsubsection{UPX}The Ultimate Packer for eXecutables\footnote{\url{https://upx.github.io/}} is a packer commonly used by malware authors to compress and obfuscate their malicious executables. When UPX packs an executable, it compresses the original executable's code, data, and resources using a compression algorithm. UPX then appends a decompression stub to the compressed data. This stub is responsible for decompressing the executable in memory when it is run. Upon execution, the operating system loads the UPX-packed binary into memory, including the decompression stub. Then, the stub decompresses the packed data back to their original form in memory, and the control is transferred to the entry point of the now-decompressed executable. In our experiments, the LZMA algorithm has been used to compress the executables.

\subsubsection{Enigma Protector}The Enigma Protector\footnote{\url{https://enigmaprotector.com/}} is a Windows software program for protecting, licensing and securing software programs. It is primarily used by software developers to safeguard their applications against illegal copying, hacking, modification, and reverse engineering. Enigma encrypts the executable files, making the code unreadable without the correct decryption key. When the protected application starts, the decryptor unpacks the original code in memory, ensuring that it runs as intended. In addition, it employs various methods of code obfuscation, mutation, and virtualization to protect the computer programs.
\subsubsection{Themida}Themida\,\footnote{\url{https://www.oreans.com/Themida.php}} is a software protector designed to prevent unauthorized use and reverse engineering of software programs. Themida uses various sophisticated techniques, including encryption, code virtualization, anti-debugging, and anti-tampering techniques (e.g., checksums, cryptographic hashes), to secure software against cracking, tampering, and other attacks.
\subsubsection{MPress}MPress\footnote{\url{https://www.autohotkey.com/mpress/mpress_web.htm}} is a packer for PE32/PE32+/.NET executable formats. Similarly to UPX, MPress packs an executable by compressing the original executable's code and data with the LZMA compression algorithm and uses an in-place decompression technique to decompress the executable without additional memory overhead. 
\subsubsection{Hyperion}Hyperion\footnote{\url{https://nullsecurity.net/tools/binary.html}} is a crypter for PE32/PE32+ files that converts a PE file into an encrypted version capable of self-decrypting at runtime. Hyperion takes a PE binary as input, loads the entire file into memory, calculates its checksum, and attaches this checksum to the file. Then, it generates a random key to encrypt the checksum and the input file using AES-128. The resulting encrypted output is then copied into the container data section. Its container serves as both a decryptor and PE loader. It loads the encrypted file into memory, decrypts it, and starts the file's execution.
\subsubsection{Amber}Amber\footnote{\url{https://github.com/EgeBalci/amber}} is a position-independent (reflective) PE loader that loads and executes the original PE file in memory. The executable files packed with Amber begin by allocating memory to load the PE file. Then, the loader reads the PE file structure and maps it into the allocated memory, making the necessary adjustments and relocations of addresses within the code. Amber then resolves any dependencies and imports that the PE file requires. 
Finally, Amber transfers control to the entry point of the PE file, starting the execution of the loaded executable directly from memory without leaving traces on the hard disk.
\subsubsection{Mangle}Mangle\footnote{\url{https://github.com/optiv/Mangle/}} is not a packer, but a tool that obfuscates well-known Indicators of Compromise (IoC) based strings and replaces them with random characters. In addition, Mangle allows the cloning of code-signing certificates of known legitimate PE files.
Code signing is a widespread security measure where the developer digitally signs an executable to verify its authenticity and integrity. The signature confirms that the file has not been altered since it was signed and helps establish trust in the software's source. In practice, the majority of legitimate software developers code-sign their executables to establish trust with users and operating systems. Conversely, malware authors typically do not have access to valid code-signing certificates for signing their malicious payloads. In our experiments, we cloned the certificate of the Microsoft Word executable file (e.g., WINWORD.EXE).
\subsubsection{Nimcrypt2}Nimcrypt2\footnote{\url{https://github.com/icyguider/Nimcrypt2}} is a free packer for PE/PE+/.NET that uses Nim-RunPE\footnote{\url{https://github.com/S3cur3Th1sSh1t/Nim-RunPE}} for reflective PE-loading from memory, and NimGetSyscallStub\,\footnote{\url{https://github.com/S3cur3Th1sSh1t/NimGetSyscallStub}} for dynamically obtaining system calls from a clean copy of the "ntdll.dll" file. 

\subsection{Dataset}
\label{sec:bodmas}
The \texttt{BODMAS} dataset~\cite{DBLP:conf/sp/YangCLA021} was collected from August 2019 and September 2020 and consists of 57,293 malware samples belonging to 581 malware families and 77,142 benign samples. The analysis with the Detect It Easy tool\footnote{\url{https://github.com/horsicq/Detect-It-Easy}} revealed the most common packers and protectors used (Cf. Table~\ref{tab:top_bodmas_packers}), showing that 35,316 out of 57,293 (59.90\%) malware and 40,243 out of 77,142 (52.17\%) benign samples are packed, highlighting that packing is slightly more prevalent in malware.

\begin{table}[ht]
\centering
\caption{Top 10 packers and protectors from the \texttt{BODMAS} dataset identified by PEiD.}
\label{tab:top_bodmas_packers}
\begin{tabular}{lclc}
\toprule
\multicolumn{2}{c}{Goodware} & \multicolumn{2}{c}{Malware}            \\ \midrule
Packer            & Count   & Packer                        & Count \\ \midrule
UPX               & 1812     & UPX                           & 9674   \\
Smart Assembly    & 1605     & Petite                        & 3795   \\
AsProtect         & 535      & ASPack                        & 1771   \\
VMProtect         & 172      & DxPack                        & 1654   \\
PyInstaller       & 139      & MPRESS                        & 1174   \\
Obsidium          & 136      & Obsidium                      & 904    \\
.NET Reactor      & 136      & tElock                        & 705    \\
Dotfuscator       & 125      & VMProtect                     & 591    \\
PE Compact        & 88       & PECompact                     & 580    \\
AsPack            & 79       & ENIGMA                        & 88     \\
\bottomrule
\end{tabular}
\end{table}

\section{Experiments}
\label{sec:experiments}
In what follows, we detail the conducted experiments to answer the research questions outlined in Section~\ref{sec:methodology}.

\subsection{Experiment I-A: malware detection - "no packed executables"}

In this experiment, we trained the malware detectors with 22,500 malicious and 22,500 benign unpacked executables randomly sampled from the \texttt{BODMAS} dataset. This subset of data has been split into training (36,000 executables - 80\%), validation (4500 executables - 10\%) and test sets (4500 executables - 10\%). In addition, we randomly sampled 4500 packed executables (2250 benign and 2250 malicious) to compare the performance of the detectors against unpacked and packed executables. The detectors have been evaluated based on their true positive rate (TPR) and true negative rate (TNR), also known as sensitivity and specificity, respectively.  Sensitivity measures how good a model is at correctly classifying malicious executables. Low sensitivity poses a cybersecurity risk, as it means that more malicious software goes undetected. Specificity indicates how good a model is at correctly classifying benign executables as benign. High specificity is also important as it ensures that benign software is not mistakenly flagged as malicious. 

\begin{figure*}[ht]
    \centering
    \includegraphics[width=\textwidth]{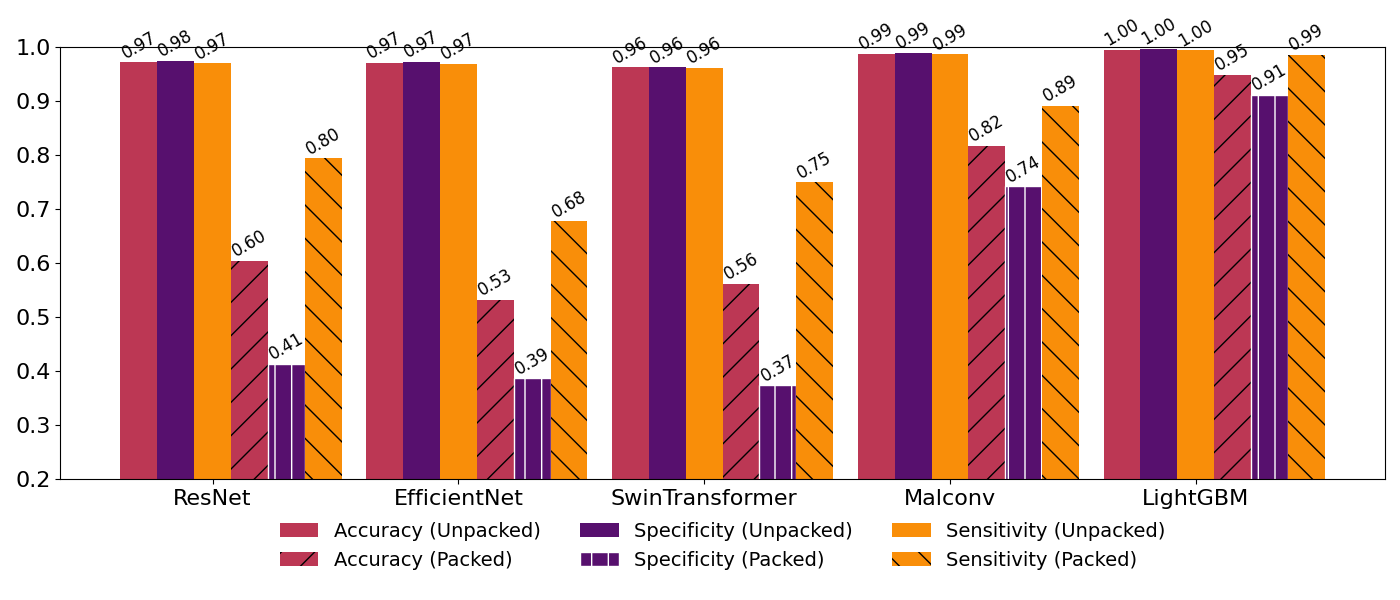}
    \caption{Experiment I-A: Performance metrics of ML-based malware detectors trained without packed executables on the "packed" and "unpacked" test sets.}
   \label{fig:boxplots_metrics_models_trained_with_unpacked_executables} 
\end{figure*}

The results in Figure~\ref{fig:boxplots_metrics_models_trained_with_unpacked_executables} show that while the specificity and sensitivity of malware detectors are moderately high for unpacked executables, they drop noticeably for packed executables. From all detectors, the LightGBM model appears to be more robust to packing. Its robustness can be attributed to the use of various features for representing a computer program, including structural and metadata features, each of which provides a different perspective on the data. Consequently, even if packing changes the statistical properties of the executables files, the LightGBM model can rely on metadata features to reach a decision. On the other hand, visual methods such as ResNet, EfficientNet, and VisionTransformer base their decision solely on the greyscale image representation of the computer program and thus, packing dramatically affects their ability to detect malware correctly, as the visual patterns these methods have learned are significantly distorted. 

\begin{table*}[ht]
\centering
\caption{Experiment I-A: Performance metrics of the detectors trained with only unpacked executables.}
\label{tab:unpacked_detectors_vs_packed_test_set}
\resizebox{\textwidth}{!}{%
\begin{tabular}{lcccccccccc}
\toprule
Detector & Metric & Unpacked test set  & UPX    & Themida & Enigma & MPress  & Hyperion & Amber & Mangle & Nimcrypt2  \\ \midrule
\multirow{2}{*}{ResNet}          & TNR    & 0.9751       & 0.7011 & 0.2931  & 0.0872 & 0.7716  & 0.4816 & 0.8947 & 0.9933 & 0.5971 \\
                                 & TPR    & 0.9707       & 0.8491 & 0.5753  & 0.9008 & 0.6123 & 0.5155 & 0.1301 & 0.2167 & 0.6221  \\ \midrule
\multirow{2}{*}{EfficientNet}    & TNR & 0.9738 & 0.6490 & 0.1659  & 0.0735 & 0.7618 & 0.4660  & 0.8321  & 0.9518 & 0.6477 \\
                                 & TPR & 0.9702 & 0.8181 & 0.8458  & 0.8325 & 0.5196 & 0.5887 & 0.0745  & 0.4717 & 0.5446  \\ \midrule
\multirow{2}{*}{SwinTransformer} & TNR & 0.9636 & 0.7134 & 0.1720  & 0.1997 & 0.868  & \textbf{0.5126} & 0.7125 & 0.9785 & 0.5925 \\
                                 & TPR & 0.9627 & 0.8328 & 0.7395  & 0.7692 & 0.2879 & 0.5555 & 0.2692 & 0.3094 & 0.5927  \\ \midrule
\multirow{2}{*}{MalConv}         & TNR    &  0.9902      & 0.6507 & 0.2946  & 0.4456 & 0.6498  & 0.0214 & 0.4578 & 0.9967 & \textbf{0.8353} \\
                                 & TPR    &  0.9871      & 0.9233 & 0.9927  & 0.9842 & 0.7946   & 0.9891 & 0.5794 & \textbf{0.7717} & 0.3605  \\ \midrule
\multirow{2}{*}{LightGBM}        & TNR & \textbf{0.9973}       & \textbf{0.7345} & \textbf{0.8127}  & \textbf{0.8270} & \textbf{0.8729}  & 0.0 & 0.2472 & \textbf{1.0} & 0.5451\\ 
                                 & TPR & \textbf{0.9951}       & \textbf{0.9310} & \textbf{0.9941}  & \textbf{0.9935} & \textbf{0.9906}  & \textbf{1.0} & \textbf{0.9100} & 0.3106 & \textbf{0.6734} \\ \bottomrule
\end{tabular}%
}
\end{table*}

The executables belonging to the packed test set in Figure~\ref{fig:boxplots_metrics_models_trained_with_unpacked_executables} have been randomly selected, so we did not have full control over how they were packed. To address this issue, we conducted an experiment wherein we tested the malware detectors against benign and malicious executables in the test set packed with the packers described in Section~\ref{sec:packers}. These packers each present unique characteristics affecting the entropy, structure, and the content of the resulting packed executable. The results, as summarised in Table~\ref{tab:unpacked_detectors_vs_packed_test_set}, reveal marked variability in detector performance when confronted with executables packed by different tools. Additionally, we have a clear trend with executables packed with sophisticated protectors such as Themida and Enigma significantly affecting the performance of end-to-end and greyscale malware detectors while packers that use encryption (e.g., Hyperion, Nimcrypt2) greatly affect the LightGBM detector. In addition, LightGBM fails to detect most malware obfuscated by Mangle. Mangle distinguishes itself from conventional packers (e.g., UPX, MPress) in its approach to obfuscation. Rather than compressing or encrypting the data, Mangle allows cloning a code-signing certificate from a legitimate PE file and obfuscates well-known IoC-based strings. To confirm which of the two manipulations performed by Mangle is more effective against the LightGBM model, we conducted an ablation study. The results in Table~\ref{tab:mangle_ablation_study} indicate that cloning a code-signing certificate significantly affects the LightGBM model as it may heavily weigh the presence of a valid code signature as an indicator of benign nature. 

\begin{table}[ht]
\centering
\caption{Mangle ablation study.}
\label{tab:mangle_ablation_study}
\begin{tabular}{lcc}
\toprule
LightGBM            & Certificate cloning + IoCs obfuscation & Certificate cloning \\ \midrule
TNR         & 1.0    &    1.0                     \\
TPR         & 0.3106 &    0.3139                    \\ \bottomrule
\end{tabular}
\end{table}

\begin{mdframed}[backgroundcolor=gray!10,linewidth=0.5pt]
    \noindent \textbf{Finding 1:} ML-based malware detectors trained without packed executables are not robust against packing. The impact of packing greatly varies depending on the type of detector. Grayscale detectors exhibit the worst performance, as packing greatly affects the structure of the executables and, consequently, the learned visual patterns. On the other hand, LightGBM faces challenges in accurately detecting benign software encrypted with Hyperion, Amber, and Nimcrypt2 and handling malware obfuscated with Mangle, primarily due to issues related to the code-signing certificate.
\end{mdframed}

\subsection{Experiment I-B: malware classification - "no packed executables".} In this experiment, we trained malware classifiers without packed executables. To this end, a random subset of 50 unpacked executables has been selected from each malware family (malware families with fewer than 50 unpacked executables have been excluded). This approach ensures that trained classifiers are trained and tested against a standardised set of data, reducing variability that could arise from different sample sizes per family. The resulting dataset of 2,650 executables from 53 families has been split into training, validation, and test sets, with the training, validation and test sets containing 40, 5, and 5 executables belonging to each family, respectively. The classifier's performance has then been evaluated against the test set, which does not include any packed executables, and against packed versions of the executables in the test set created using the packers described in Section~\ref{sec:packers}. 


\begin{table*}[ht]
\centering
\caption{Experiment I-B: Accuracy of the "unpacked" classifiers against different test sets.}
\label{tab:unpacked_classifiers_vs_packed_test_set}
\resizebox{\textwidth}{!}{%
\begin{tabular}{lccccccccc}
\toprule
Detector & Unpacked test set   & UPX    & Themida & Enigma & MPress & Hyperion & Amber & Mangle & Nimcrypt2\\ \midrule
ResNet   & 0.8113              & 0.1832 & 0.0617  & 0.0535 & 0.1547 & \textbf{0.0518} & 0.0244 & 0.4048 & 0.0346\\
EfficientNet    & 0.7283 & 0.1145 & 0.0535  & 0.0412 & 0.1208 & 0.0279 & \textbf{0.0634} & 0.3333 & 0.0269\\
SwinTransformer & 0.7925 & 0.1298 & 0.0576  & 0.0576 & 0.1623 & 0.0239 & 0.0 & 0.3905 & \textbf{0.0423}\\ \midrule
MalConv  & 0.8717              & 0.2214 & 0.0700  & 0.0947 & 0.2189 & 0.0319 & 0.0293 & \textbf{0.8333} & 0.0269 \\
LightGBM & \textbf{0.90567}             & \textbf{0.2672} & \textbf{0.2099}  & \textbf{0.1893} & \textbf{0.3245} & 0.0398 & 0.0195 & 0.7619 & 0.0192 \\ \bottomrule
\end{tabular}%
}
\end{table*}

Table~\ref{tab:unpacked_classifiers_vs_packed_test_set} presents a detailed breakdown of the accuracy rates achieved by the different classifiers 
across the various packed and unpacked test scenarios. It can be observed that while the classifiers performed well in unpacked executables, with accuracies ranging from about 80\% to 90\%, their performance dramatically declined to approximately 20\% when tested against packed executables.
This decline is attributed to the substantial structural changes applied by packing, which were not represented in the training data. Packed executables differ significantly from their unpacked counterparts, challenging the classifiers' generalisation ability.

\begin{mdframed}[backgroundcolor=gray!10,linewidth=0.5pt]
    \noindent \textbf{Finding 2:} Similar to ML-based malware detectors, malware classifiers trained without packed executables struggle to classify packed malware into their respective families correctly. This suggests that packed executables have to be used during training.
\end{mdframed}

\subsection{Experiment II: malware detection - "only packed goodware or packed malware".} 
\begin{figure*}[ht] 
  \centering 
  \begin{subfigure}[b]{0.45\textwidth}
    \includegraphics[width=\linewidth]{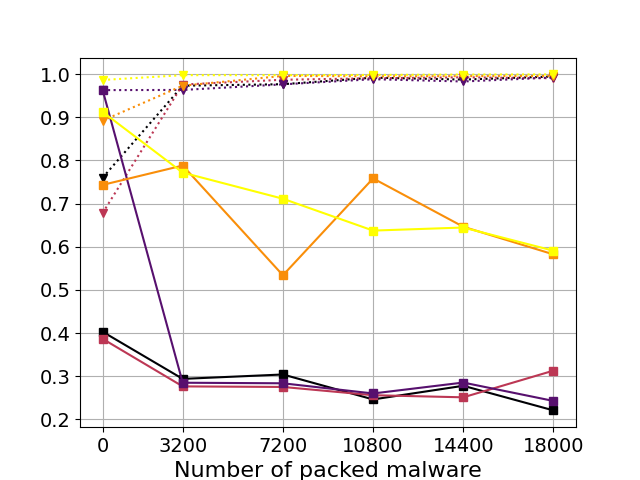}
  \end{subfigure}
  \hfill 
  \begin{subfigure}[b]{0.45\textwidth}
    \includegraphics[width=\linewidth]{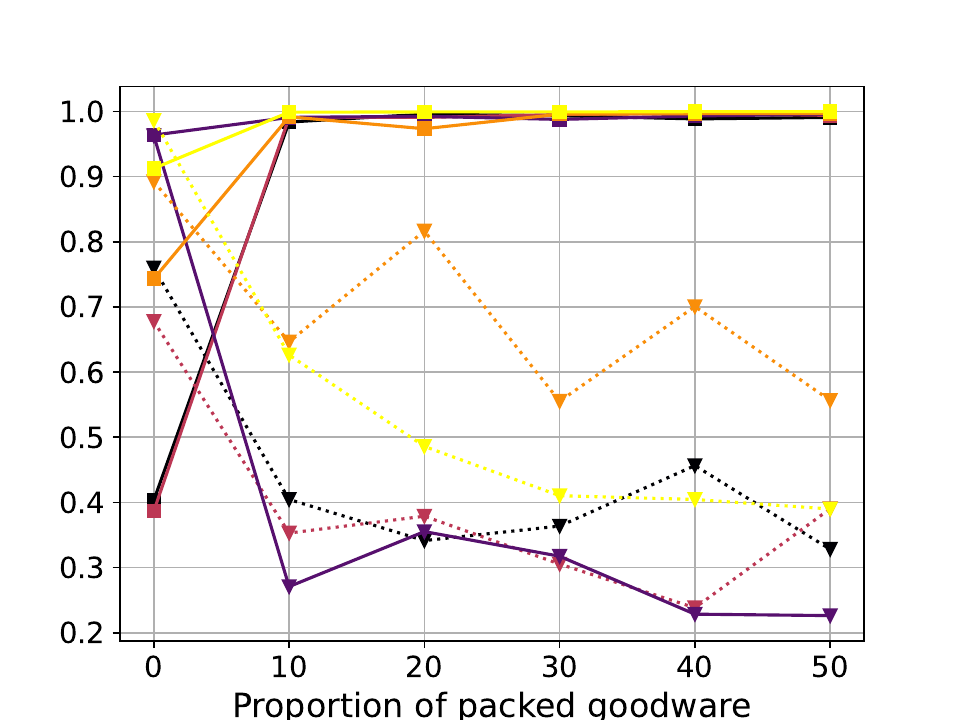}
  \end{subfigure}
  \caption{Experiment II-A: Performance metrics of the ML-based malware detectors trained with packed malware or packed goodware. Squares denote Specificity and triangles Sensitivity. \fillbox{ResNetCol} denotes ResNet, \fillbox{EfficientNetCol} denotes EfficientNet, \fillbox{SwinTransformerCol} denotes SwinTransformer,\fillbox{MalConvCol} denotes MalConv, and \fillbox{LightGBMCol} denotes LightGBM, } 
  \label{fig:lineplot_mlw_detectors_packed_malware_or_packed_goodware} 
\end{figure*}

Using as baseline the dataset described in \textbf{Experiment I-A}, 22,500 additional packed malware has been added to the training (+18,000 packed malware), validation (+2250 packed malware), and test sets (+2250 packed malware). The resulting training set consists of up to 18,000 packed malware, 18,000 unpacked malware and 18,000 unpacked goodware. 
In this experiment, the ML-based malware detectors have been trained using varying quantities of packed malicious executables, e.g., +3600, +7200, +10800, +14400 and +18000, to determine whether the exposure to packed malicious executables during training leads to a predisposition of the detectors to classify packed executables, i.e., benign and malicious, as malicious. Figure~\ref{fig:lineplot_mlw_detectors_packed_malware_or_packed_goodware} illustrates the performance metrics of the malware detectors as the number of packed malware seen during training increases. The results show that as the number of packed malware seen during training increases, the true positive rate of the detectors increases. In contrast, the true negative rate decreases, implying that the detectors are more likely to classify packed goodware as malware, i.e., interpreting packing as a sign of maliciousness. Similarly, when the number of packed goodware seen during training increases, the specificity of the detectors increases while the TPR decreases, inducing a bias toward associating packing with benign intent.

\begin{mdframed}[backgroundcolor=gray!10,linewidth=0.5pt]
    \noindent \textbf{Finding 3:} Excluding packed benign software from the training set biases the detector towards interpreting packing as a sign of maliciousness. Conversely, excluding packed malware biases the detector towards interpreting packing as an indication of benign intent. This indicates that both packed and unpacked executables should be included during training.
\end{mdframed}

\subsection{Experiment III: good versus bad packers.} 

In this experiment, we trained the detectors using a dataset with goodware packed by UPX, Themida, Enigma, and MPress, and malware packed with Hyperion, Amber, Mangle, and Nimcrypt2.
We then tested the detectors on executables packed with the previously mentioned packers. The results in Table~\ref{tab:good_vs_bad_packers} reveal that the detectors are strongly inclined to classify all executables packed with the "good" packers as benign and those with the "bad" packers as malicious. This shows a significant bias, increasing the risk of false positives and negatives based on the packers predominantly used for packing benign and malicious executables in the training set.


\begin{table*}[ht]
\centering
\caption{Experiment III-A: Performance metrics of malware detectors trained with goodware packed with UPX, Themida, Enigma, and MPress, and malware packed with Hyperion, Amber, Mangle, and Nimcrypt2. }
\label{tab:good_vs_bad_packers}
\resizebox{\textwidth}{!}{%
\begin{tabular}{lccccccccc}
\toprule
Detector & Metric      & UPX  & Themida & Enigma & MPress & Hyperion & Amber & Mangle & Nimcrypt2  \\ \midrule
\multirow{2}{*}{ResNet}   & TNR & \textbf{0.9943} & \textbf{0.9736}  & \textbf{0.9986} & \textbf{0.9827} & 0.2466 & 0.0251 & 0.4740 & 0.0101 \\
                          & TPR & 0.3483 & 0.0105  & 0.1769 & 0.0700 & \textbf{0.9964} & \textbf{0.9959} & \textbf{0.9711} & \textbf{1.0} \\ \midrule
\multirow{2}{*}{EfficientNet}   & TNR & \textbf{0.7940} & \textbf{0.9538}  & \textbf{0.9978} & \textbf{0.8009} & \textbf{0.7068} & 0.2158 & 0.5112 & 0.0478 \\
                                & TPR & 0.1707 & 0.0259  & 0.0137 & 0.3106 & 0.5369 & \textbf{0.9255} & \textbf{0.7794} & \textbf{0.9938} \\ \midrule
\multirow{2}{*}{SwinTransformer}   & TNR & \textbf{0.9967} & \textbf{0.9873}  & \textbf{0.9993} & \textbf{0.9876} & 0.3165 & 0.0275 & 0.2923 & 0.0078 \\
                                   & TPR & 0.4026 & 0.1078  & 0.1941 & 0.0557 & \textbf{0.9932} & \textbf{0.9935} & \textbf{0.9833} & \textbf{1.0} \\ \midrule
\multirow{2}{*}{MalConv}  & TNR & \textbf{1.0}    & \textbf{0.9915}  & \textbf{1.0} & \textbf{0.992}  & 0.0 & 0.0 & 0.0949 & 0.0 \\
                          & TPR & 0.0397 & 0.0053  & 0.0 & 0.0512 & \textbf{1.0} & \textbf{1.0} & \textbf{0.9994} & \textbf{1.0} \\ \midrule
\multirow{2}{*}{LightGBM} & TNR & \textbf{1.0} & \textbf{1.0}     & \textbf{1.0} & \textbf{0.9947} & 0.0 & 0.0057 & 0.5880 & 0.0 \\ 
                          & TPR & 0.0 & 0.0010  & 0.0 & 0.0348 & \textbf{1.0} & \textbf{0.9992} & \textbf{0.9989} & \textbf{1.0} \\ \bottomrule
\end{tabular}%
}
\end{table*}

\begin{mdframed}[backgroundcolor=gray!10,linewidth=0.5pt]
    \noindent \textbf{Finding 4:} A lack of overlap between packers used in goodware and malware will bias the detectors to distinguish between benign software and malicious software based on the packing routines rather than the actual content. This indicates that the detector must be trained with a diverse set of packed and unpacked executables from both categories to ensure it learns to identify malware based on intrinsic characteristics rather than superficial packing methods.
\end{mdframed}

\subsection{Experiment IV: packers detection.}
This experiment shifts focus from training a malware detector to developing a packer detector. The primary objective is to evaluate the feasibility of using machine learning to classify executables based on the packer employed. The packer detectors have been trained using the dataset from \textbf{Experiment I-A} obfuscated with the packers defined in Section~\ref{sec:packers}. The results in Table~\ref{tab:packers_detectors} show that all detectors achieved over 99\% accuracy, indicating that the structural manipulations applied by different packers are easily detectable and classifiable.

\begin{table*}[ht]
\centering
\caption{Experiment IV: Accuracy of the packer detectors against different test sets scenarios.}
\label{tab:packers_detectors}
\resizebox{\textwidth}{!}{%
\begin{tabular}{p{1.25in}p{.75in}lccccccccc}
\toprule
Packer detector & Test set  & UPX    & Themida & Enigma & MPress  & Hyperion & Amber & Mangle & Nimcrypt2  \\ \midrule
ResNet           & 0.9918      & 0.9837 & 0.9998 & 0.9987 & 0.9702 & 0.9982 & 0.9994 & 0.9882 & \textbf{1.0}\\  
EfficientNet     & 0.9946 & 0.9891 & 0.9970 & 0.9986 & 0.9811 & 0.9985 & 0.9997 & 0.9951 & \textbf{1.0}\\  
SwinTransformer  & 0.9925 & 0.9724 & 0.9978 & 0.9989 & 0.9878 & 0.9941 & 0.9994 & 0.9849 & \textbf{1.0}\\  \midrule
MalConv  & 0.9945      & 0.9946 & 0.9925 & 0.9921 & 0.9840 & \textbf{1.0} & \textbf{1.0} & 0.9964 & \textbf{1.0}\\ 
LightGBM & \textbf{0.99985}     & \textbf{1.0} & \textbf{1.0}  & \textbf{1.0} & \textbf{0.9991} & \textbf{1.0} & \textbf{1.0} & \textbf{1.0} & \textbf{1.0}\\ \bottomrule

\end{tabular}%
}
\end{table*}


\begin{mdframed}[backgroundcolor=gray!10,linewidth=0.5pt]
    \noindent \textbf{Finding 5:} The structural manipulations applied by different packers can be easily detected and classified by grayscale-based, byte-based, feature-based classifiers with no effort. 
\end{mdframed}

\subsection{Experiment V-A: malware detection - "unpacked and packed executables".} 
\begin{figure}[ht]
    \centering
    \includegraphics[width=.6\columnwidth]{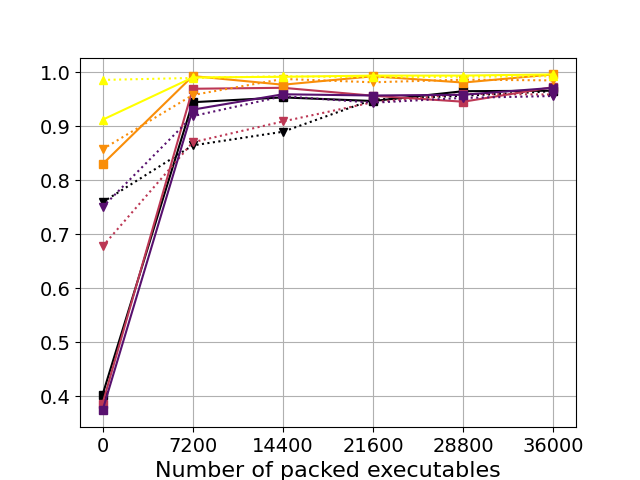}
    \caption{Experiment V-A: Performance metrics of the ML-based malware detectors trained with packed and unpacked goodware and malware on the packed test set. Uses the same colour and shape mapping as Figure \ref{fig:lineplot_mlw_detectors_packed_malware_or_packed_goodware}.}
    \label{fig:lineplots_models_with_packed_and_unpacked_examples}
\end{figure}

In this experiment, the malware detectors were trained with varying amounts of packed goodware and malware to determine the impact of including packed executables during training. The results in Figure~\ref{fig:lineplots_models_with_packed_and_unpacked_examples} reveal that the inclusion of packed executables during the training phase significantly improves the detectors' performance, especially when exposed to 7,200 packed executables (3,600 goodware and 3,600 malware).
\begin{table*}[ht]
\centering
\caption{Experiment V-A: Performance metrics of malware detectors trained with both packed and unpacked executables. }
\label{tab:detectors_vs_packed_test_set}
\resizebox{\textwidth}{!}{%
\begin{tabular}{lcccccccccc}
\toprule
Detector & Metric & Unpacked test set & UPX  & Themida & Enigma & MPress & Hyperion & Amber & Mangle & Nimcrypt2  \\ \midrule
\multirow{2}{*}{ResNet}          & TNR & 0.972       & 0.5822 & 0.4865  & 0.7332 & 0.6822 & 0.3709 & 0.7149 & 0.99 & 0.4374 \\
                                 & TPR & 0.9604      & 0.9353 & 0.474  & 0.4687 & 0.7589  & 0.7188 & 0.2766 & 0.2027 & 0.8155 \\ \midrule
\multirow{2}{*}{EfficientNet}    & TNR & 0.9729 & 0.6205 & 0.5749  & 0.7282 & 0.7338 & \textbf{0.5981} & \textbf{0.8700} & 0.9871 & 0.4531 \\
                                 & TPR & 0.9591 & 0.9293 & 0.6178  & 0.3789 & 0.7215 & 0.6074 & 0.0597 & \textbf{0.4778} & 0.7638 \\ \midrule
\multirow{2}{*}{SwinTransformer} & TNR & 0.9716 & 0.6173 & 0.4218  & 0.5739 & 0.7196 & 0.2971 & 0.7519 & 0.9790 & 0.4623 \\
                                 & TPR & 0.9564 & 0.8784 & 0.6724  & 0.6362 & 0.6123 & 0.7384 & 0.2128 & 0.4256 & 0.7210 \\ \midrule
\multirow{2}{*}{MalConv}         & TNR &  0.992      & \textbf{0.9039} & \textbf{0.9852}  & \textbf{0.9841} & 0.8609 & 0.1981 & 0.6290 & 0.9981 & \textbf{1.0} \\
                                 & TPR &  0.9849     & 0.8362 & 0.2143  & 0.4881 & 0.6907  & 0.9568 & 0.5172 & 0.4172 & 0.0 \\ \midrule
\multirow{2}{*}{LightGBM} & TNR & \textbf{0.9960}      & 0.8827 & 0.8830  & 0.8529 & \textbf{0.9391} & 0.0214 & 0.4364 & \textbf{1.0} & 0.3086 \\ 
 & TPR & \textbf{0.9951}      & \textbf{0.9621} & \textbf{0.9873}  & \textbf{0.9065} & \textbf{0.9777} & \textbf{0.9995} & \textbf{0.9935} & 0.1028 & \textbf{0.8338} \\ \bottomrule
\end{tabular}%
}
\end{table*}

Next, the malware detectors were trained with equal numbers of unpacked and packed goodware and malware (18.000 packed goodware, 18.000 unpacked goodware, 18000 packed malware, 18.000 packed malware) to investigate the impact of packing on detectors trained with balanced proportions of benign and malicious executables. 
Table~\ref{tab:detectors_vs_packed_test_set} shows that the detectors' performance varies significantly across different packers, suggesting that some packers are more challenging for certain detectors.
While all models perform well against unpacked executables, packing introduces an additional layer of complexity, reducing their effectiveness. Packing can be used for evasion purposes, with specific packers proving effective at evading detection and others leading to a higher number of false negatives, resulting in legitimate software being mistakenly identified as malicious. For instance, MalConv struggles with benign executables packed with Hyperion and Amber, ResNet has issues with all packers beyond Mangle, and LightGBM experiences challenges with goodware packed with Hyperion, Amber, and Nimcrypt2. Additionally, malware packed with Themida, Engima, Amber, and Mangle are hardly detected by MalConv, while LightGBM is easily evaded by Mangle-packed malware.


\subsection{Experiment V-B: Train a malware classifier with packed and unpacked executables.} 
This experiment extends the dataset from \textbf{Experiment I-B} with packed executables from each family where available\footnote{Some malware families do not have any packed executables. Packed executables from different malware families may have been packed with different versions of the same packer, e.g., UPX.}. Each family has been extended with 50 packed executables, 40 for training and 5 for validation and testing. The results in Table~\ref{tab:classifiers_vs_packed_test_set} show that the resulting classifiers exhibit very low accuracy across different packers, highlighting the challenge of accurately classifying packed malware.

In contrast, recent studies utilising datasets like the \texttt{Malimg} dataset~\cite{10.1145/2016904.2016908} or the Microsoft Malware Classification Challenge dataset~\cite{DBLP:journals/corr/abs-1802-10135} report significantly higher accuracy rates. These datasets typically contain a single family that comprises all the packed executables, simplifying the classification task, and resulting in highly overoptimistic performance. This stark difference in performance underscores the importance of building newer, more complex datasets that better reflect real-world scenarios with malware packed in various ways across multiple families.

\begin{table*}[ht]
\centering
\caption{Experiment V-B: Accuracy of the malware classifiers trained with both packed and unpacked executables against different packed test sets.}
\label{tab:classifiers_vs_packed_test_set}
\resizebox{\textwidth}{!}{%
\begin{tabular}{lccccccccc}
\toprule
Detector & Unpacked test set   & UPX    & Themida & Enigma & MPress & Hyperion & Amber & Mangle & Nimcrypt2\\ \midrule
ResNet           & 0.7585 & 0.1679 & 0.0658  & 0.0494 & 0.1434  & 0.0398 & 0.0 & 0.4476 & 0.0192 \\
EfficientNet     & 0.6491 & 0.1832 & 0.0741  & 0.0576 & 0.1321  & 0.0159 & 0.0 & 0.3762 & 0.0385 \\
SwinTransformer  & 0.8151 & 0.1985 & 0.0494  & 0.0329 & 0.1774  & 0.0120 & 0.0098  & 0.4238 & \textbf{0.0538} \\ \midrule
MalConv          & 0.8755 & 0.2366 & 0.0493  & 0.1152 & 0.2037  & \textbf{0.0359} & \textbf{0.0195} & \textbf{0.8286} & 0.0038 \\
LightGBM         & \textbf{0.9019} & \textbf{0.2671} & \textbf{0.1523}  & \textbf{0.1111} & \textbf{0.3019}  & \textbf{0.0359} & 0.0146 & 0.7619 & 0.0038 \\ \bottomrule
\end{tabular}%
}
\end{table*}

\begin{mdframed}[backgroundcolor=gray!10,linewidth=0.5pt]
    \noindent \textbf{Finding 6:} Packing limits the effectiveness of ML-based malware detectors and classifiers even when the same ratio of packed and unpacked executables is used during training.
\end{mdframed}

\subsection{Experiment VI-A: "single packer" detector.} 
\label{sec:single_packer_detector}

This experiment simulates a scenario where a detector must determine the maliciousness of executables packed with a packer encountered during training. For each packer, we extended the training set from \textbf{Experiment I-A} with benign (+4,000) and malicious (+4,000) executables obfuscated with that packer. Then, the detectors are evaluated against executables obfuscated with the same packer. The results in Table~\ref{tab:single_packer_detection} show that while the detectors' performance generally improved compared to those trained with packed executables randomly selected from the \texttt{BODMAS} dataset, the consistency varied between packers and detectors. For instance, detectors struggled against executables packed with Hyperion, Amber, and Nimcrypt2.

\begin{table*}[ht]
\centering
\caption{Experiment VI-A: Performance metrics of malware detectors trained and tested with executables packed using the same packer.}
\label{tab:single_packer_detection}
\begin{tabular}{lccccccccc}
\toprule
                              &        & \multicolumn{2}{c}{UPX} & \multicolumn{2}{c}{Themida} & \multicolumn{2}{c}{Enigma} & \multicolumn{2}{c}{MPress}  \\ \cmidrule{3-10}
Detector                      & Metric & Unpacked    & Packed    & Unpacked      & Packed      & Unpacked      & Packed     & Unpacked      & Packed   \\ \midrule
\multirow{2}{*}{ResNet}       & TNR    & 0.9738      & 0.9731    & 0.9578        & 0.8621      & 0.9564        & 0.9027     & 0.976         & 0.9507        \\
                              & TPR    & 0.9613      & 0.9474    & 0.9573        & 0.9237      & 0.9684        & 0.9461     & 0.9596        & 0.9634    \\
\multirow{2}{*}{EfficientNet} & TNR    & 0.9742      & 0.9625    & 0.9609        & 0.8509      & 0.9647        & 0.9647     & 0.9644        & 0.9524 \\
                              & TPR    & 0.9658      & 0.9466    & 0.964         & 0.9384      & 0.9418        & 0.9418     & 0.96          & 0.9639     \\
\multirow{2}{*}{SwinT}        & TNR    & 0.9533      & 0.9577    & 0.9587        & 0.8758      & 0.9604        & 0.9545     & 0.972         & 0.9618    \\
                              & TPR    & 0.964       & 0.9379    & 0.964         & 0.9286      & 0.9716        & 0.9424     & 0.9476        & 0.9630  \\ \midrule
\multirow{2}{*}{MalConv}      & TNR    & 0.988       & \textbf{0.9910}    & 0.9849        & 0.9679      & \textbf{0.9947}        & 0.9776     & 0.9778        & 0.9702    \\
                              & TPR    & 0.9867      & 0.9690    & 0.9876        & 0.9692      & 0.9867        & 0.9705     & 0.9889        & 0.9857     \\
\multirow{2}{*}{LightGBM}     & TNR    & \textbf{0.9978}      & 0.9894    & \textbf{0.9973}        & \textbf{0.9893}      & \textbf{0.9947}        & \textbf{0.9834}     & \textbf{0.9947}        & \textbf{0.9871}    \\
                              & TPR    & \textbf{0.9942}      & \textbf{0.9922}    & \textbf{0.9951}        & \textbf{0.9936}      & \textbf{0.9938}        & \textbf{0.9935}     & \textbf{0.9924}        & \textbf{0.9906}   \\ \midrule \vspace{1pt} \\ \midrule

                              &        & \multicolumn{2}{c}{Hyperion} & \multicolumn{2}{c}{Amber} & \multicolumn{2}{c}{Mangle} & \multicolumn{2}{c}{Nimcrypt2}  \\ \cmidrule{3-10}
Detector                      & Metric & Unpacked    & Packed    & Unpacked      & Packed      & Unpacked      & Packed     & Unpacked      & Packed   \\ \midrule
\multirow{2}{*}{ResNet}       & TNR    & 0.96           & 0.7359      & 0.9769       & 0.9545     & 0.9818        & 0.9824     & 0.9662         & 0.4384  \\
                              & TPR    & 0.9516         & 0.9263      & 0.968        & 0.8944     & 0.9707        & 0.9767     & 0.9498         & 0.9274    \\
\multirow{2}{*}{EfficientNet} & TNR    & 0.9693         & 0.4388      & 0.9724       & \textbf{0.9801}     & 0.9711        & 0.9781     & 0.9676         & 0.3997    \\
                              & TPR    & 0.9431         & 0.9813      & 0.9653       & 0.8331     & 0.9716        & 0.9839     & 0.9676         & 0.9599   \\
\multirow{2}{*}{SwinT}        & TNR    & 0.9773         & 0.6601      & 0.9796       & 0.9412     & 0.9627        & 0.9733     & 0.9617         & 0.4581      \\
                              & TPR    & 0.9524         & 0.9140      & 0.9476       & 0.8625     & 0.9564        & 0.9683     & 0.9413         & 0.9389  \\ \midrule
\multirow{2}{*}{MalConv}      & TNR    & 0.9716         & 0.4136      & 0.9618       & 0.8904     & 0.9542        & 0.9866     & 0.964          & 0.9048    \\
                              & TPR    & \textbf{0.9889}      & \textbf{0.9986}      & 0.9929       & 0.8085     & 0.9876        & 0.99       & \textbf{0.9933}       & \textbf{0.9964}   \\
\multirow{2}{*}{LightGBM}     & TNR    & \textbf{0.9782}         & \textbf{0.9961}      & \textbf{0.9969}       & 0.9658     & \textbf{0.9973}        & \textbf{0.9938}     & \textbf{0.9747}         & \textbf{1.0}  \\
                              & TPR    & 0.9867         & 0.0118      & \textbf{0.9951}       & \textbf{0.8993}     & \textbf{0.9942}        & \textbf{0.9917}     & 0.9844         & 0.0 \\ \bottomrule 
\end{tabular}%
\end{table*}

\subsection{Experiment VI-B: "single packer" classifier.}  

This experiment assesses the classifiers' ability to generalise across different malware families when using a known packer to pack unseen executables. We utilised a subset of samples from the \texttt{BODMAS} dataset, selecting 250 unpacked executables from each family. Families with fewer than 250 unpacked executables were excluded to ensure uniformity and data sufficiency. The resulting dataset consists of 17 malware families. Each family's 250 unpacked executables were divided into training, validation, and test sets of 200, 25, and 25 executables, respectively. We trained multiple classifiers using a combined dataset of unpacked and packed executables with a single packer. Thus, each classifier was exposed to both unpacked and packed malware samples obfuscated with a specific packer during training. These classifiers were then evaluated on a test set packed with the same packer seen during the training.
This evaluation helps us observe the classifier's performance in a controlled environment where the packing method remains constant across the training and testing phases. The results in Table~\ref{tab:single_packer_classifier} illustrate that augmenting the training set with packed executables does not necessarily improve the accuracy at test time against executables packed with the packer used to augment the training set. More precisely, while all classifiers maintain high accuracy when trained and tested with executables packed with UPX, Themida, and Enigma, they struggle against executables packed with MPress, Hyperion, and Nimcrypt2.

\begin{table*}[ht]
\centering
\caption{Experiment VI-B: Accuracy of the malware classifiers trained and tested with executables packed with a single packer.}
\label{tab:single_packer_classifier}
\begin{tabular}{lcccccccc}
\toprule
                                & \multicolumn{2}{c}{UPX} & \multicolumn{2}{c}{Themida} & \multicolumn{2}{c}{Enigma} & \multicolumn{2}{c}{MPress}  \\ \cmidrule{2-9}
Detector                        & Unpacked    & Packed    & Unpacked      & Packed      & Unpacked      & Packed     & Unpacked      & Packed   \\ \midrule
\multirow{2}{*}{ResNet}         & 0.8376      & 0.7396    & 0.84          & 0.8040      & 0.8259        & 0.8351     & 0.84          & 0.3608  \\
\multirow{2}{*}{EfficientNet}   & 0.7976      & 0.5384    & 0.7906        & 0.6023      & 0.7741        & 0.6702     & 0.8376        & 0.2901  \\
\multirow{2}{*}{SwinT}          & 0.8282      & 0.6923    & 0.8071        & 0.8069      & 0.8424        & 0.8586     & 0.7976        & 0.3325  \\ \midrule
\multirow{2}{*}{MalConv}        & 0.9106      & \textbf{0.9112}    & \textbf{0.9529}        & 0.8934      & 0.9224        & 0.8848     & 0.9529        & 0.3915  \\
\multirow{2}{*}{LightGBM}       & \textbf{0.9459}      & 0.8757    & 0.9435        &  \textbf{0.9365}           & \textbf{0.9482}        & \textbf{0.9110}     & \textbf{0.9553}        & \textbf{0.4292}  \\ \midrule \vspace{1pt} \\ \midrule

                             & \multicolumn{2}{c}{Hyperion} & \multicolumn{2}{c}{Amber} & \multicolumn{2}{c}{Mangle} & \multicolumn{2}{c}{Nimcrypt2}  \\ \cmidrule{2-9}
Detector                     & Unpacked    & Packed    & Unpacked      & Packed      & Unpacked      & Packed     & Unpacked      & Packed   \\ \midrule
\multirow{2}{*}{ResNet}         & 0.8071         & \textbf{0.7854}      & 0.8306       & \textbf{0.9309}     & 0.8353        & 0.8942     & 0.7835         & 0.4447 \\
\multirow{2}{*}{EfficientNet}   & 0.8541         & 0.7193      & 0.8118       & 0.8980     & 0.8306        & 0.8978     & 0.8023         & 0.3953 \\
\multirow{2}{*}{SwinT}          & 0.8376         & 0.7005      & 0.8212       & 0.9013     & 0.8305        & 0.9088     & 0.8259         & \textbf{0.5341} \\ \midrule
\multirow{2}{*}{MalConv}        & 0.9412         & 0.5495      & 0.9412       & 0.5        & 0.9482        & \textbf{0.9416}     & 0.9294         & 0.3153 \\
\multirow{2}{*}{LightGBM}       & \textbf{0.9482}         & 0.4764      & \textbf{0.9529}       & 0.9243     & \textbf{0.9506}        & 0.9343     & \textbf{0.9483}         & 0.0635 \\ \bottomrule 
\end{tabular}%
\end{table*}

\begin{mdframed}[backgroundcolor=gray!10,linewidth=0.5pt]
    \noindent \textbf{Finding 7:} Experiments VI-A and VI-B suggest that while exposure to packed executables can improve the detector's and classifier's robustness, it is not guaranteed, as the effectiveness is highly dependent on the used packing method.
\end{mdframed}

\subsection{Experiment VII-A: "withheld" packer detector.} 
\label{sec:witheld_packer_detector}

\begin{table*}[ht]
\centering
\caption{Experiment VII-A: Performance metrics of malware detectors trained with executables packed with all packers except one and evaluated against executables packed using the "withheld" packer.}
\label{tab:leave_one_detection}
\begin{tabular}{lccccccccc}
\toprule
                              &        & \multicolumn{2}{c}{UPX} & \multicolumn{2}{c}{Themida} & \multicolumn{2}{c}{Enigma} & \multicolumn{2}{c}{MPress}  \\ \cmidrule{3-10}
Detector                      & Metric & Unpacked    & Packed    & Unpacked      & Packed      & Unpacked      & Packed     & Unpacked      & Packed   \\ \midrule
\multirow{2}{*}{ResNet}       & TNR    & 0.9316      & 0.6799    & 0.9702        & 0.8076      & 0.9609        & 0.2473     & 0.9729        & 0.9062\\
                              & TPR    & 0.972       & 0.8431    & 0.9547        & 0.3557      & 0.9702        & 0.7541     & 0.956         & 0.4786\\
\multirow{2}{*}{EfficientNet} & TNR    & 0.9627      & 0.7842    & 0.972         & 0.5746      & 0.9738        & 0.5552     & 0.9609        & 0.8911\\
                              & TPR    & 0.9671      & 0.8121    & 0.9653        & 0.5421      & 0.9702        & 0.6341     & 0.9542        & 0.5312\\
\multirow{2}{*}{SwinT}        & TNR    & 0.9618      & 0.8493    & 0.9427        & 0.9150      & 0.9538        & 0.4579     & 0.9738        & 0.8858\\
                              & TPR    & 0.9591      & 0.7612    & 0.9631        & 0.3293      & 0.964         & 0.8327     & 0.9578        & 0.3859\\ \midrule
\multirow{2}{*}{MalConv}      & TNR    & \textbf{0.9867}      & \textbf{0.9503}    & \textbf{0.9818}        & 0.902       & \textbf{0.9942}        & \textbf{0.9913}     & \textbf{0.9867}        & 0.7547\\
                              & TPR    & 0.9822      & 0.8828    & \textbf{0.9862}        & 0.9193      & \textbf{0.984}         & 0.7081     & 0.9787        & 0.9693\\
\multirow{2}{*}{LightGBM}     & TNR    & 0.9733      & 0.9292    & 0.9707        & \textbf{0.9481}      & 0.9698        & 0.9575     & 0.9689        & \textbf{0.9444}\\
                              & TPR    & \textbf{0.9871}      & \textbf{0.925}     & 0.9849        & \textbf{0.9795}      & 0.9827        & \textbf{0.9267}     & \textbf{0.9827}        & \textbf{0.9701}\\ \midrule \vspace{1pt} \\ \midrule

                              &        & \multicolumn{2}{c}{Hyperion} & \multicolumn{2}{c}{Amber} & \multicolumn{2}{c}{Mangle} & \multicolumn{2}{c}{Nimcrypt2}  \\ \cmidrule{3-10}
Detector                      & Metric & Unpacked    & Packed    & Unpacked      & Packed      & Unpacked      & Packed     & Unpacked      & Packed   \\ \midrule
\multirow{2}{*}{ResNet}       & TNR    & 0.9644         & \textbf{0.8563}      & 0.9747       & 0.9668     & 0.9618        & 0.9758     & 0.9756         & 0.6099\\
                              & TPR    & 0.9573         & 0.3553      & 0.9507       & 0.0417     & 0.9698        & \textbf{0.4878}     & 0.9591         & 0.6020\\
\multirow{2}{*}{EfficientNet} & TNR    & 0.9702         & 0.6252      & 0.9533       & 0.8961     & 0.9724        & 0.9834     & 0.9613         & 0.4972 \\
                              & TPR    &0.9542         & 0.6201      & 0.9693       & 0.0638     & 0.9676        & 0.3033     & 0.972          & \textbf{0.7019}\\
\multirow{2}{*}{SwinT}        & TNR    &0.9662         & 0.7689      & 0.9604       & \textbf{0.9696}     & 0.9667        & 0.9833     & 0.9609         & 0.5322\\
                              & TPR    &0.9596         & 0.3121      & 0.9684       & 0.0352     & 0.9618        & 0.1961     & 0.9662         & 0.6471\\ \midrule
\multirow{2}{*}{MalConv}      & TNR     & \textbf{0.9964}         & 0.1204      & \textbf{0.9804}       & 0.8297     & \textbf{0.9756}        & 0.9928     & \textbf{0.9956}         & \textbf{1.0}\\
                              & TPR    & 0.9818         & 0.9268      & 0.9796       & 0.3175     & 0.988         & 0.4622     & 0.9782         & 0.0\\
\multirow{2}{*}{LightGBM}     & TNR    & 0.9742         & 0.2408      & 0.9738       & 0.5094     & 0.9711        & \textbf{0.9952}     & 0.9951         & 0.7222\\
                              & TPR   & \textbf{0.9862}         & \textbf{0.9899}      & \textbf{0.9818}       & \textbf{0.9861}     & \textbf{0.9853}        & 0.2283     & \textbf{0.9938}         & 0.4612 \\ \bottomrule 
\end{tabular}%
\end{table*}

This experiment simulates a scenario where malware detectors must identify executables obfuscated with a packer not encountered during training. To achieve this, the training set from Experiment I has been extended with executables obfuscated with each packer except for one, which was withheld (4,000 benign and 4,000 malicious executables packed with each packer). The resulting detectors were then evaluated against executables obfuscated with the packer excluded from training. As shown in Table~\ref{tab:leave_one_detection}, the performance of the detectors sharply drops depending on the withheld packer, highlighting the limitations of ML-based approaches in detecting executables obfuscated with unknown packers. This study utilised only a subset of eight well-known packers, yet in the real world, there are numerous available packers, posing a substantial challenge to the robustness and generalisation capabilities of such malware detection systems.
This underscores the need to complement static ML-based malware detectors with more resilient detection methods capable of adapting to the vast and evolving landscape of packing techniques.

\subsection{Experiment VII-B: "withheld" packer classifier.} 

Using the subset introduced in Experiment VI-B, we train multiple classifiers with unpacked executables and packed executables obfuscated with all packers except one, which was withheld. The resulting classifiers were then evaluated against executables obfuscated by the packer excluded from training. This setup assesses the classifiers' ability to generalise across different malware families when an unknown packer is used to pack the executables at test time. The results in Table~\ref{tab:withheld_packer_classifier} demonstrate that the classifier's accuracy significantly declines.

\begin{table*}[ht]
\centering
\caption{Experiment VII-B: Accuracy of malware classifiers trained and tested with executables packed with all packers except one.}
\label{tab:withheld_packer_classifier}
\begin{tabular}{lcccccccc}
\toprule
                                & \multicolumn{2}{c}{UPX} & \multicolumn{2}{c}{Themida} & \multicolumn{2}{c}{Enigma} & \multicolumn{2}{c}{MPress}  \\ \cmidrule{2-9}
Detector                        & Unpacked    & Packed    & Unpacked      & Packed      & Unpacked      & Packed     & Unpacked      & Packed   \\ \midrule
\multirow{2}{*}{ResNet}         & 0.8212      & 0.4201    & 0.8353        & 0.1037      & 0.8612        & 0.2251     & 0.8424        & 0.3184\\
\multirow{2}{*}{EfficientNet}   & 0.8235      & 0.3669    & 0.8118        & 0.0893      & 0.8118        & 0.2435     & 0.8424        & 0.3349\\
\multirow{2}{*}{SwinT}    & 0.8565      & 0.4260    & 0.8565        & 0.1902      & 0.8424        & 0.1518     & 0.8235        & 0.3585       \\ \midrule
\multirow{2}{*}{MalConv}  & 0.9436      & \textbf{0.5385}    & 0.9459        & \textbf{0.2767}      & \textbf{0.9459}        & \textbf{0.4372}     & 0.9435        & 0.3962      \\
\multirow{2}{*}{LightGBM}   & \textbf{0.9576}      & 0.3728    & \textbf{0.9482}        & 0.2421      & \textbf{0.9459}        & 0.2644     & \textbf{0.9529}        & \textbf{0.4882}    \\ \midrule \vspace{1pt} \\ \midrule

                             & \multicolumn{2}{c}{Hyperion} & \multicolumn{2}{c}{Amber} & \multicolumn{2}{c}{Mangle} & \multicolumn{2}{c}{Nimcrypt2}  \\ \cmidrule{2-9}
Detector                     & Unpacked    & Packed    & Unpacked      & Packed      & Unpacked      & Packed     & Unpacked      & Packed   \\ \midrule
\multirow{2}{*}{ResNet}   & 0.84           & \textbf{0.1156}      & 0.8329       & 0.0493     & 0.8565        & 0.5474     & 0.8494         & 0.1176        \\
\multirow{2}{*}{EfficientNet}   & 0.84           & \textbf{0.1156}      & 0.8329       & 0.0493     & 0.8565        & 0.5474     & 0.8494         & 0.1176 \\
\multirow{2}{*}{SwinT}    & 0.8353         & 0.0967      & 0.8706       & \textbf{0.1053}     & 0.8541        & 0.4781     & 0.84           & 0.12          \\ \midrule
\multirow{2}{*}{MalConv}   & 0.9435         & 0.1061      & 0.9388       & 0.0461     & 0.9459        & \textbf{0.9088}     & 0.9318         & 0.1106      \\
\multirow{2}{*}{LightGBM}   & \textbf{0.9459}         & 0.0613      & \textbf{0.96}         & 0.0559     & \textbf{0.9506}        & 0.6788     & \textbf{0.9529}         & 0.0824     \\ \bottomrule 
\end{tabular}%
\end{table*}

\begin{mdframed}[backgroundcolor=gray!10,linewidth=0.5pt]
    \noindent \textbf{Finding 7:} Experiments VII-A and VII-B highlight the difficulties of ML-based malware detectors and classifiers to correctly classify packed executables even when obfuscated executables are included in the training set, as the variety and sophistication of the packing techniques make it difficult for the models to generalize to unseen packers. This significant drop in accuracy for packed executables highlights the necessity for integrating more advanced detection techniques, such as dynamic analysis, for dealing with packed malware. 
\end{mdframed}

\subsection{Experiment IX: anti-malware engines in the industry versus packers}
The packed test set used in Experiments~\ref{sec:single_packer_detector} and~\ref{sec:witheld_packer_detector} was submitted to VirusTotal to evaluate the performance of eight anti-malware products that are described as static machine learning-based malware detectors on the VirusTotal blog post. We focused only on these engines since the classifiers that this work considers are also static. Results in Table~\ref{tab:antimalware_engines_results} show that anti-malware engines in the industry have learned to associate packing with maliciousness, i.e., they label packed executables as malicious, with some exceptions such as the executables packed with Nimcrypt2 which are mostly classified as benign by Engine 1 and Engine 4. These results have significant implications because packing, as shown in Section~\ref{sec:bodmas}, is not only used by malicious actors, but it is also a legitimate technique used to reduce the size of the executables and to protect intellectual property rights. As a result, labeling packed executables as malicious will generate a high rate of false positives (benign executables incorrectly classified as malicious), which in turn can overwhelm security teams with alerts and consume unnecessary resources analyzing these files. In addition, security researchers use VirusTotal to gather threat intelligence and as a source of labeled data for training their own machine learning models to detect malware. If almost every packed executable is labeled as malicious, the training data will be polluted, leading to models that are biased and inaccurate. It is well understood that some packers might have been flagged as malicious, as they are used by many malware; nonetheless, packers like UPX, MPress, and Themida are used by numerous benign applications for binary protection. On the contrary, samples using Nimcrypt2 were all treated as benign, signifying that the engine has training issues. Finally, the fact that two engines failed to analyse a significant number of samples implies that there are significant issues in parsing some files. The latter is rather alarming as it can be easily abused by threat actors.  

\begin{mdframed}[backgroundcolor=gray!10,linewidth=0.5pt]
    \noindent \textbf{Finding 8:} Anti-malware engines are biased toward packing, associating it with malicious intent.
\end{mdframed}

\begin{table*}[ht]
\centering
\caption{Accuracy of eight anti-malware engines against packed goodware and malware. A dash (-) indicates that the anti-malware engine failed to analyze the packed samples.}
\label{tab:antimalware_engines_results}
\begin{tabular}{cccccccccc}
\toprule
Packer                                       & Subset     & Engine 1           & Engine 2      & Engine 3                    & Engine 4                & Engine 5                     & Engine 6                   & Engine 7                    & Engine 8               \\ \midrule
\multicolumn{1}{c}{\multirow{2}{*}{UPX}}     & Goodware & 0.8884                     & -                          & 0.5147                     & -                     & 0.9748                     & 0.9845                     & 0.9528                     & 0.9397                     \\
\multicolumn{1}{c}{}                         & Malware  & 0.9007                     & -                          & 0.9233                     & -                     & 0.8789                     & 0.8354                     & 0.9338                     & 0.8267                     \\
\multicolumn{1}{c}{\multirow{2}{*}{Themida}} & Goodware & 0.4443                     & 0.1669                     & 0.0041                     & 0.3908                     & 0.5903                     & 0.4860                     & 0.1985                     & 0.2814                     \\
\multicolumn{1}{c}{}                         & Malware  & 0.9364                     & 0.9804                     & 0.9839                     & 0.8537                     & 0.9462                     & 0.9565                     & 0.9927                     & 0.9892                     \\
\multicolumn{1}{c}{\multirow{2}{*}{Enigma}}  & Goodware & 0.1370                     & 0.0570                     & 0.0014                     & 0.1420                     & 0.0404                     & 0.2242                     & 0.0324                     & 0.0101                     \\
\multicolumn{1}{c}{}                         & Malware  & 0.9720                     & 0.9950                     & 0.9885                     & 0.9669                     & 0.9878                     & 0.9547                     & 0.9914                     & 0.9935                     \\
\multirow{2}{*}{MPress}                      & Goodware & 0.7829                     & 0.7321                     & 0.5350                     & 0.6955                     & 0.7994                     & 0.9443                     & 0.7775                     & 0.6001                     \\
                                             & Malware  & 0.9756                     & 0.9842                     & 0.9799                     & 0.9391                     & 0.9892                     & 0.9621                     & 0.9971                     & 0.9914                     \\
\multirow{2}{*}{Hyperion}                    & Goodware & 0.4214                     & 0.0019                     & 0.0                        & 0.0                        & 0.0582                     & 0.0117                     & 0.0                        & 0.0                        \\
                                             & Malware  & 0.9011                     & 0.9362                     & 0.9303                     & 0.9116                     & 0.9212                     & 0.9362                     & 0.9326                     & 0.9280                     \\
\multirow{2}{*}{Amber}                       & Goodware & 0.3496 & 0.0024 & 0.0299 & 0.4369 & 0.0    & 0.2462 & 0.0    & 0.0    \\
                                             & Malware  & 0.9680 & 1.0    & 1.0    & 0.5777 & 0.9967 & 0.6318 & 0.9812 & 0.9877 \\
\multirow{2}{*}{Mangle}                      & Goodware & 0.9966 & -      & 0.9890 & 0.9215 & 0.9957 & 0.9995 & 0.9880 & 0.9909 \\
                                             & Malware  & 0.8650 & -      & 0.9933 & 0.9453 & 0.9978 & 0.9967 & 0.9955 & 0.9944 \\
\multirow{2}{*}{Nimcrypt2}                   & Goodware & 0.9982 & -      & 0.3422 & 1.0    & 0.3914 & 0.1624 & 0.0    & 0.0    \\
                                             & Malware  & 0.0    & -      & 0.7558 & 0.0    & 0.6934 & 0.9911 & 0.9772 & 0.9742 \\ \bottomrule
\end{tabular}
\end{table*}

\section{Conclusions}
The recent advances in machine learning and artificial intelligence have enabled a plethora of applications and facilitated many tasks and automation. Undeniably, they have both boosted the quality of cybersecurity measures; however, this is not a panacea. Our work demonstrates the pitfalls that well-established models in the field of malware detection face when dealing with obfuscated malware. Our dataset analysis reveals the underlying similarities of the samples and justifies, to a great extent, the visual similarities and why visual methods exhibit such excellent results. For instance, having very similar files (at the byte level) or files sharing identical sections can obviously lead to visually similar images. By identifying these inherent biases, we performed several targeted and extensive experiments exhibiting the strong dependence of all these models on training with obfuscated samples. Indeed, the efficacy of all models significantly drops when they face specific packers. 

What is interesting to note is that our work demonstrates a realistic, low barrier technique to evade ML-based malware classifiers. A real-world adversary will require non-trivial technical skill to create adversarial samples to bypass an ML-based model. Most malware authors are not experts in machine learning, and the required changes to bypass detection and maintain original functionality are not trivial and not always achievable. Nevertheless, using another packer to protect the malware is an obvious strategy for the adversary in this scenario, and, as we show, depending on the packer and model pair, can be very successful. 

To this end, we argue that while such detection models can be very efficient in specific cases, they are not as efficient as claimed. More importantly, on many occasions, they simply detect the packer, which is a completely different task. Of course, some packers are known to be used heavily by malware, so identifying their presence is a strong signal that the file is malicious. 

Finally, our work demonstrates the importance of having datasets considering temporal factors and evasion strategies. New samples will most likely use new methods and packers and, as a result, will have different features. By considering the temporal evolution of the samples, one can craft better features that reflect these changes. 
Moreover, it is crucial to evaluate malware detectors not only for their ability to adapt to evolving malware but also for their robustness against adversarial attacks and packing techniques. Adversarial attacks, where an attacker deliberately manipulates the malware to evade detection, significantly degrades the performance of the detectors. Similarly, packing techniques obscure the malicious code, making it more challenging to detect. 
Therefore, malware detectors must be continuously trained to keep the models up to date with the new emerging features and rigorously evaluated against evasion strategies to ensure that the malware detectors remain effective in real-world conditions, beyond the sterile laboratory environment.     

\section*{Acknowledgements}
This project has received funding from Enterprise Ireland and the European Union’s Horizon 2020 Research and Innovation Programme under the Marie Skłodowska-Curie grant agreement No 847402 and was supported by the European Commission under the Horizon Europe Programme, as part of the projects, SafeHorizon (Grant Agreement no. 101168562), CyberSecPro (\url{https://www.cybersecpro-project.eu}) (Grant Agreement no. 101083594) and LAZARUS (\url{https://lazarus-he.eu/}) (Grant Agreement no. 101070303).

The content of this article does not reflect the official opinion of the European Union. Responsibility for the information and views expressed therein lies entirely with the authors.

We would like to thank Cormac Doherty and UCD's Centre for Cybersecurity and Cybercrime Investigation for their support.

\bibliographystyle{unsrt}  
\bibliography{references}

\end{document}